\documentclass[sigconf,natbib,screen]{acmart}

\usepackage{booktabs} 
\usepackage{adjustbox}
\usepackage{algorithm, algpseudocode}
\usepackage{amsmath}
\usepackage{graphics}
\usepackage{epsfig}
\usepackage{graphicx}
\usepackage{xcolor}
\usepackage[skip=0pt]{caption}
\usepackage{subfigure}
\usepackage{balance}
\usepackage{multirow}
\usepackage{mathrsfs}
\usepackage{acronym}
\usepackage{placeins}
\usepackage{xcolor}
\usepackage[inline]{enumitem}

\usepackage{url}
\usepackage{xspace}
\usepackage{bbm}

\newcommand\BibTeX{B{\sc ib}\TeX}
\newcommand{\kld}{\mathit{KL}}
\newcommand{\vmf}{\text{vMF}}

\AtBeginDocument{%
  \providecommand\BibTeX{{%
    \normalfont B\kern-0.5em{\scshape i\kern-0.25em b}\kern-0.8em\TeX}}}

\acrodef{CF}{collaborative filtering}
\acrodef{MIAs}{membership inference attacks}
\acrodef{MIA}{membership inference attack}
\acrodef{MF}{matrix factorization}
\acrodef{LFM}{latent factor model}
\acrodef{NCF}{neural collaborative filtering}
\acrodef{DL-MIA}{Debiasing Learning for Membership Inference Attacks against recommender systems}
\acrodef{VAE}{variational auto-encoder}
\acrodef{ItemBase}{Item-based collaborative filtering}
\acrodef{MLP}{multilayer perceptron}
\acrodef{vMF}{von Mises Fisher}

\newcommand{\header}[1]{\vspace*{1mm}\noindent\textbf{#1.}}

\definecolor{french_blue}{RGB}{0, 112, 182}

\allowdisplaybreaks

\copyrightyear{2022}
\acmYear{2022}
\setcopyright{acmlicensed}
\acmConference[KDD '22]{Proceedings of the 28th ACM SIGKDD Conference on Knowledge Discovery and Data Mining}{August 14--18, 2022}{Washington, DC, USA}
\acmBooktitle{Proceedings of the 28th ACM SIGKDD Conference on Knowledge Discovery and Data Mining (KDD '22), August 14--18, 2022, Washington, DC, USA}
\acmPrice{15.00}
\acmDOI{10.1145/3534678.3539392}
\acmISBN{978-1-4503-9385-0/22/08}

\begin{document}

\author{Zihan Wang}
\authornote{These authors contributed equally to this work.}
\affiliation{
 \institution{Shandong University 
 \city{}
 \country{}}
}
\email{zihanwang.sdu@gmail.com}

\author{Na Huang}
\authornotemark[1]
\affiliation{%
 \institution{Shandong University 
 \city{}
 \country{}}
}
\email{hn.z@mail.sdu.edu.cn}

\author{Fei Sun}
\affiliation{%
 \institution{Alibaba Group 
 \city{}
 \country{}}
}
\email{ofey.sunfei@gmail.com}

\author{Pengjie Ren}
\affiliation{%
 \institution{Shandong University 
 \city{}
 \country{}}
}
\email{jay.ren@outlook.com}

\author{Zhumin Chen}
\affiliation{%
 \institution{Shandong University 
 \city{}
 \country{}}
}
\email{chenzhumin@sdu.edu.cn}

\author{Hengliang Luo}
\affiliation{%
 \institution{Meituan 
 \city{}
 \country{}}
}
\email{luohengliang@meituan.com}

\author{Maarten de Rijke}
\affiliation{%
 \institution{University of Amsterdam 
 \city{}
 \country{}}
}
\email{m.derijke@uva.nl}

\author{Zhaochun Ren}
\authornote{Corresponding author.}
\affiliation{
 \institution{Shandong University 
 \city{}
 \country{}}
}
\email{zhaochun.ren@sdu.edu.cn}

\fancyfoot{}
\renewcommand{\shortauthors}{Zihan Wang et al.}

\title[Debiasing Learning for Membership Inference Attacks Against Recommender Systems]{Debiasing Learning for Membership Inference Attacks \\ Against Recommender Systems}


\begin{abstract}
Learned recommender systems may inadvertently leak information about their training data, leading to privacy violations.
We investigate privacy threats faced by recommender systems through the lens of membership inference.
In such attacks, an adversary aims to infer whether a user's data is used to train the target recommender.
To achieve this, previous work has used a shadow recommender to derive training data for the attack model, and then predicts the membership by calculating difference vectors between users' historical interactions and recommended items.
State-of-the-art methods face two challenging problems: 
\begin{enumerate*}[label=(\roman*)]  
\item training data for the attack model is biased due to the gap between shadow and target recommenders, and 
\item hidden states in recommenders are not observational, resulting in inaccurate estimations of difference vectors.
\end{enumerate*}

To address the above limitations, we propose a \acfi{DL-MIA} framework that has four main components:
\begin{enumerate*}[label=(\roman*)]
    \item a difference vector generator,
    \item a disentangled encoder,
    \item a weight estimator, and
    \item an attack model.
\end{enumerate*}
To mitigate the gap between recommenders, a \acf{VAE} based disentangled encoder is devised to identify recommender invariant and specific features.
To reduce the estimation bias, we design a weight estimator, assigning a truth-level score for each difference vector to indicate estimation accuracy.
We evaluate \ac{DL-MIA} against both general recommenders and sequential recommenders on three real-world datasets.
Experimental results show that \ac{DL-MIA} effectively alleviates training and estimation biases simultaneously, and achieves state-of-the-art attack performance.
\end{abstract}



\maketitle

\acresetall


\section{Introduction}
\label{sec:Introduction}
The success of today's recommender systems is largely attributed to the increased availability of large-scale training data on users' private information (e.g., browsing and purchase history).
Unfortunately, various studies show that recommender systems are vulnerable to attacks, leading to the leakage of their training data and severe privacy problems~\citep{DBLP:conf/sp/ShokriSSS17,DBLP:conf/uss/Carlini0EKS19}.

In this paper, we study privacy threats faced by recommender systems through the lens of membership inference~\citep{DBLP:conf/sp/ShokriSSS17}.
Specifically, \acp{MIA} against recommender systems enable the adversary to infer whether a user's data is used to train the target recommender~\citep{DBLP:conf/ccs/ZhangRWRCHZ21}. 
The main reason for the feasibility of \ac{MIA} is overfitting, since the learned model tends to perform better on the training data~\citep{DBLP:conf/ccs/ChenYZF20}.
Revealing the membership may cause serious harm, and leak sensitive information about specific individuals, such as shopping preferences, social relationships, and location information~\citep{DBLP:conf/icml/Choquette-ChooT21}.

Existing \ac{MIA} methods show promising performance in various domains, ranging from biomedical data~\citep{DBLP:conf/ccs/0001BHM16, DBLP:conf/ndss/Hagestedt0HBT0019,homer2008resolving} to mobility traces~\citep{DBLP:conf/ndss/PyrgelisTC18}.
Despite the success, previous \ac{MIA} methods~\citep{DBLP:conf/sp/ShokriSSS17,DBLP:conf/ndss/Salem0HBF019,DBLP:conf/csfw/YeomGFJ18, DBLP:conf/sp/NasrSH19, DBLP:conf/icml/Choquette-ChooT21,DBLP:conf/ccs/LiZ21} cannot be directly applied to recommender systems, since they either require knowledge of the target model or use the predicted confidence scores of the classifier.
In \acp{MIA} against recommender systems, the target recommenders are considered inaccessible, and only recommended items, rather than confidence scores, are observational to the adversary~\citep{DBLP:conf/ccs/ZhangRWRCHZ21}.
In fact, this setting is prevalent in real-world scenarios.

\begin{figure}
  \centering
  \subfigure[Training data bias.]{
    \includegraphics[clip,trim=10mm 5mm 0mm 0mm,width=0.485\linewidth]{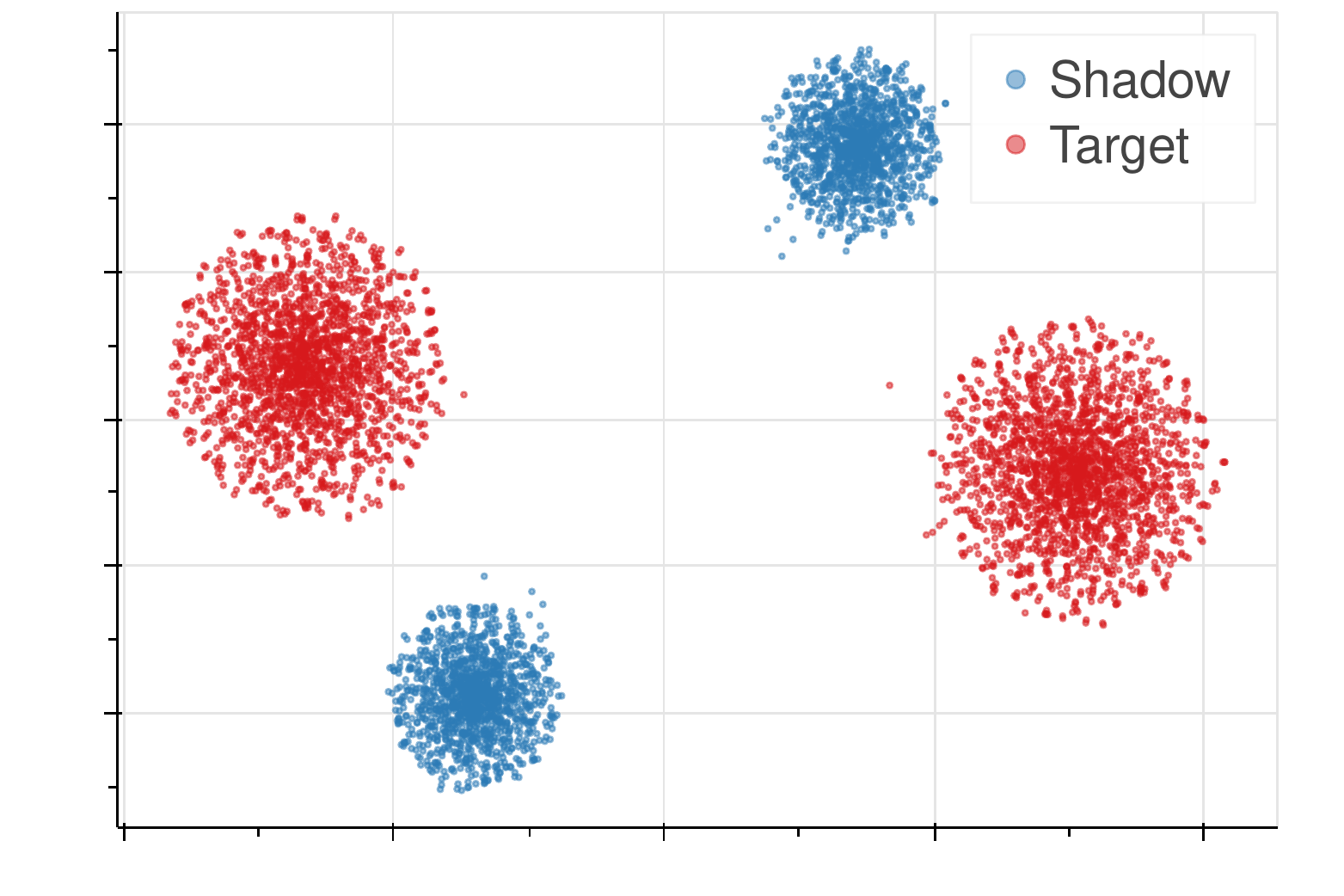}\label{fig:Training data bias.}}
  \subfigure[Estimation bias.]{\includegraphics[clip,trim=10mm 5mm 0mm 0mm,width=0.485\linewidth]{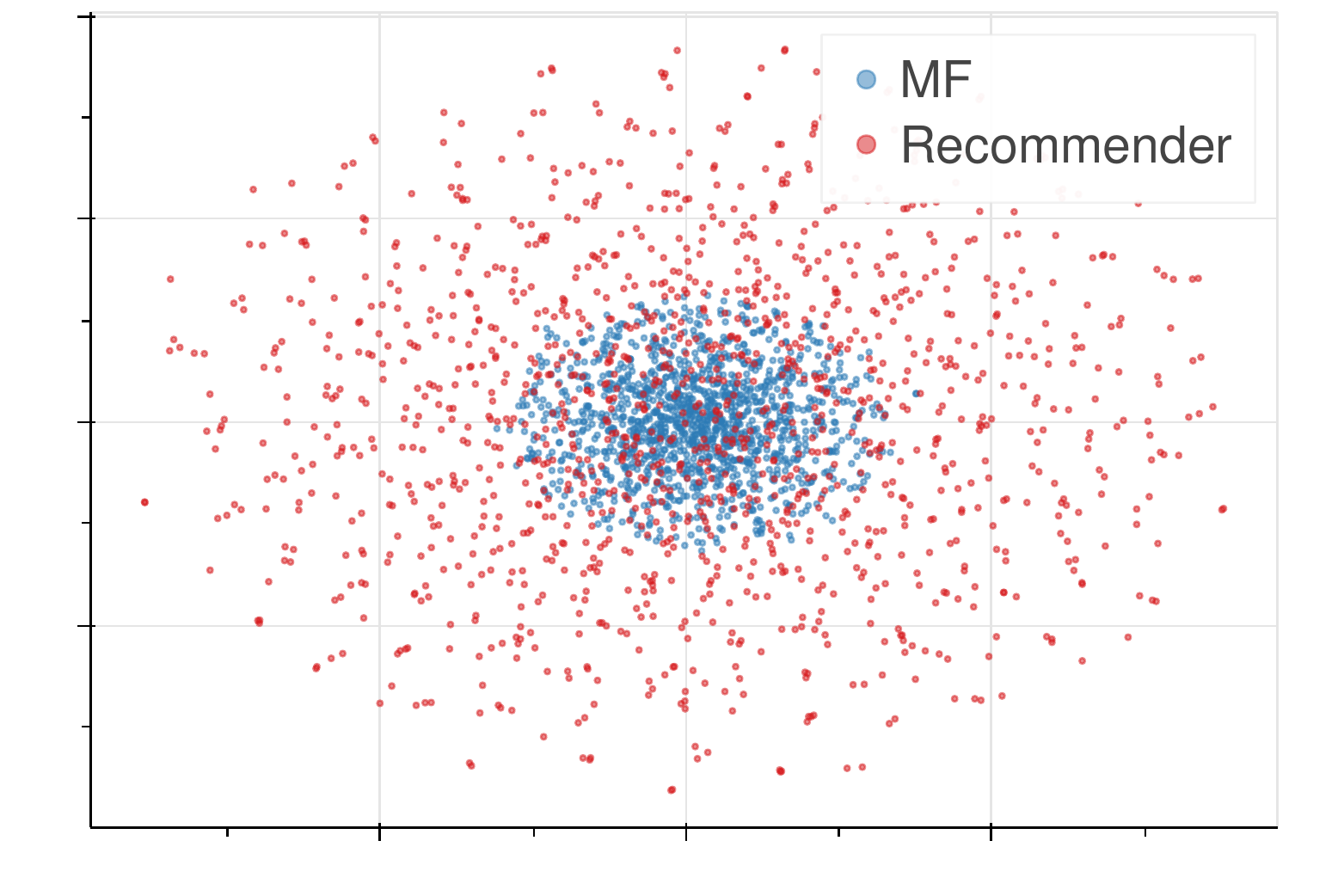}\label{fig:Estimation bias.}}
    \caption{Visualization results for the training data and estimation biases. (a) The bias between the \ac{MIA} datasets generated by the shadow recommender (blue) and target recommender (red). (b) The bias between difference vectors generated using \ac{MF} (blue) and hidden states in the recommender (red).}
  \label{fig:Motivation}
  \vspace*{-.5\baselineskip}
\end{figure}

In recent work, \citet{DBLP:conf/ccs/ZhangRWRCHZ21} infer the membership of the target recommender based on the similarity between users' historical interactions and recommended items. 
The key idea here is, for users in the training set, their historical interactions tend to be more similar to output items of the recommender.
Specifically, a shadow recommender is first established to simulate the target recommender and generate training data for the attack model.
Then, difference vectors between users' historical interactions and recommended items are computed by factorizing the user-item rating matrix.
On this basis, the attack model is able to predict the membership using difference vectors. 
This framework faces two challenging problems:
\begin{enumerate}[leftmargin=*]
\item {} \textbf{Training data for the attack model is biased.}
As mentioned above, the algorithm and dataset used by the target recommender are inaccessible~\citep{DBLP:conf/ccs/ZhangRWRCHZ21}.
In that case, the adversary may construct a shadow recommender in a completely distinct manner, resulting in a biased training dataset for the attack model.  
In Figure~\ref{fig:Training data bias.}, feature vectors from the \ac{MIA} datasets generated by target and shadow recommenders (that use different methods) are visualized by the t-SNE algorithm~\citep{JMLR:v9:vandermaaten08a}, respectively.
And there exist huge differences between the distributions of features obtained from the shadow recommender (blue) and target recommender (red).
Besides, as mentioned in~\citep{DBLP:conf/ccs/ZhangRWRCHZ21}, the attack performance drops dramatically when target and shadow recommenders use different algorithms and datasets.

To mitigate the gap between recommender systems, we employ a \ac{VAE} based encoder to disentangle features, and model recommender invariant and specific characteristics using two different distribution families. 

\item {} \textbf{The estimations of difference vectors are inaccurate.}
In this attack, as explained above, the hidden states (e.g., user and item representations) in the target recommender are not available to the adversary. 
As a result, difference vectors between user historical interactions and recommended items may be estimated inaccurately for the target recommender, leading to incorrect membership predictions.
For example, as demonstrated in Figure~\ref{fig:Estimation bias.},
the difference vectors generated by the target recommender (red) and \ac{MF} (blue) are divergently distributed.

To reduce the influence of the estimation bias, we develop a weight estimator, and learn a truth-level score for each difference vector to indicate the estimation accuracy during training.
\end{enumerate}
To address the above problems, we propose a framework, named \acfi{DL-MIA}, to simultaneously mitigate training data and estimation biases.
As illustrated in Figure~\ref{fig:An overview.}, \ac{DL-MIA} has four main components:
\begin{enumerate*}[label=(\roman*)]
    \item a difference vector generator,
    \item a disentangled encoder,
    \item a weight estimator, and
    \item an attack model.
\end{enumerate*}
During training, to simulate behavior of the target model, a shadow recommender is first constructed and learned on a shadow dataset.
Then, the generator represents users' history interactions and recommended items by factorizing the user-item rating matrix, and calculates difference vectors.
To mitigate the training data bias caused by the gap between target and shadow recommenders, the disentangled encoder is developed, and a \acf{VAE} based on two distribution families is employed to identify recommender invariant and specific features.
Next, to reduce the influence of the estimation bias, we establish a weight estimator, and assign a truth-level score for each difference vector.
Finally, the disentangled and re-weighted difference vectors, as well as membership labels, are input for the \ac{MLP} based attack model.
In addition, to facilitate the model parameter update and weight learning, an alternating training strategy is applied among the disentangled encoder, weight estimator, and attack model.

Our contributions can be summarized as follows:
\begin{enumerate*}[label=(\roman*)]
    \item To the best of our knowledge, ours is the first work to study debiasing learning for membership inference attacks against recommender systems.
    \item We develop a \ac{VAE} based disentangled encoder to mitigate training data bias caused by the gap between shadow and target recommenders.
    \item We introduce truth-level scores, and learn the weight estimator with the alternating training strategy to alleviate the estimation bias of difference vectors.
    \item Experimental results show that \ac{DL-MIA} achieves the state-of-the-art attack performance against both general and sequential recommender systems.
\end{enumerate*}



\section{Related work}
\label{sec:Related works}
We survey related work along three dimensions: 
\begin{enumerate*}[label=(\roman*)]
\item membership inference attacks,
\item general and sequential recommenders, and
\item debiasing learning.
\end{enumerate*}

\subsection{Membership inference attacks}
Recently, \acf{MIAs} have achieved pro\-mising performance in various domains, such as biomedical data~\citep{DBLP:conf/ccs/0001BHM16, DBLP:conf/ndss/Hagestedt0HBT0019,homer2008resolving} and mobility traces~\citep{DBLP:conf/ndss/PyrgelisTC18}. 
The goal of membership inference attacks is to infer the membership of individual training samples for a target model.
\citet{DBLP:conf/sp/ShokriSSS17} specify the first membership inference attack against machine learning models.
The authors propose a general formulation of membership inference attack against machine learning models, and train multiple shadow models to simulate the target model's behavior.
In that case, the training sets for multiple attack models (one for each class) are generated.
\citet{DBLP:conf/ndss/Salem0HBF019} further relax several key assumptions from~\citep{DBLP:conf/sp/ShokriSSS17}, including knowledge of the target model architecture and target dataset distribution.
\citet{DBLP:conf/csfw/YeomGFJ18} explore the relationship between attack performance and overfitting, and propose the first decision-based attack.
\citet{DBLP:conf/sp/NasrSH19} study membership inference attacks in both black-box and white-box settings.
Instead of using output scores, several recent membership attacks~\citep{DBLP:conf/icml/Choquette-ChooT21,DBLP:conf/ccs/LiZ21} assume only predicted hard labels of models are exposed, and demonstrate that label-only exposures are also vulnerable to membership leakage.
In addition,~\citet{DBLP:conf/ccs/ZhangRWRCHZ21} investigate \ac{MIA} against recommender systems, leveraging the differences between user history behaviors and output items from recommenders.

To mitigate the attacks, some defense mechanisms, including model stacking~\citep{DBLP:conf/ndss/Salem0HBF019}, dropout~\citep{DBLP:conf/ndss/Salem0HBF019}, adversarial training~\citep{DBLP:conf/ccs/NasrSH18}, differential privacy~\citep{DBLP:conf/icml/Choquette-ChooT21,DBLP:conf/ccs/LiZ21}, regularization~\citep{DBLP:conf/icml/Choquette-ChooT21,DBLP:conf/ccs/LiZ21}, and jointly
maximizing privacy and prediction accuracy~\citep{DBLP:conf/ccs/JiaSBZG19}, have been proposed. 
To protect membership privacy of recommender systems,~\citet{DBLP:conf/ccs/ZhangRWRCHZ21} design a defense mechanism, named \emph{Popularity Randomization}, and randomly recommend popular items to non-member users.

\subsection{General and sequential recommenders}
A generic recommender system aims to model users' preferences from their historical behavior.
Early attempts on recommender systems, including \acf{MF}~\citep{DBLP:reference/sp/KorenB15,DBLP:journals/computer/KorenBV09,mnih2007probabilistic,DBLP:conf/sac/PolatD05} and item-based neighborhood methods~\citep{DBLP:journals/internet/LindenSY03,DBLP:conf/kdd/KabburNK13,DBLP:conf/kdd/Koren08,DBLP:conf/www/SarwarKKR01}, typically apply \acf{CF} on users' interaction histories.
Recently, deep learning has been used to improve the performance of recommender systems by incorporating with auxiliary information~\citep{DBLP:conf/recsys/KimPOLY16, DBLP:conf/icdm/KangFWM17}, or replacing the conventional matrix factorization~\citep{he2017neural,DBLP:conf/www/SedhainMSX15}.

None of the above methods considers the order in users' behaviors or is designed for sequential recommendation.
The earliest work on sequential recommendation, FPMC~\citep{DBLP:conf/www/RendleFS10}, utilizes Markov chain to capture the transition in behavior sequences.
To further enhance the capability of modeling complex behavior, deep learning based models~\citep{DBLP:journals/corr/HidasiKBT15, DBLP:conf/aaai/ZhouMFPBZZG19,DBLP:conf/icdm/KangM18, DBLP:conf/cikm/SunLWPLOJ19,DBLP:conf/wsdm/TangW18} are devised, including recurrent neural network based~\citep{DBLP:journals/corr/HidasiKBT15, DBLP:conf/aaai/ZhouMFPBZZG19}, and attention based~\citep{DBLP:conf/icdm/KangM18, DBLP:conf/cikm/SunLWPLOJ19} methods.

\subsection{Debiasing learning}
Bias is a critical issue in modern machine learning since trained models often fail to identify the proper representations for the target predictions~\citep{chu2021learning}.
To tackle the limitation, a large number of methods have been proposed to eliminate the biases.
Specifically, to address selection bias~\citep{DBLP:conf/uai/MarlinZRS07} in datasets, propensity score~\citep{DBLP:conf/icml/SchnabelSSCJ16,DBLP:conf/wsdm/0003ZS021}, ATOP~\citep{DBLP:conf/kdd/Steck10}, and data imputation~\citep{DBLP:conf/sigir/Saito20} are utilized.
Besides, debiasing strategies such as rebalancing~\citep{DBLP:conf/sigmod/AsudehJS019}, adversarial learning~\citep{DBLP:conf/wsdm/BeigiMGAN020}, and causal modeling~\citep{DBLP:conf/nips/KusnerLRS17} are proposed to mitigate
unfairness~\citep{DBLP:conf/recsys/LinSMB19} caused by algorithm and unbalanced data.
In survey~\citep{DBLP:journals/corr/abs-2010-03240}, seven types of biases with their definitions and characteristics are summarized and introduced in detail. 
However, the work listed does not consider the biases in \ac{MIA} against recommender systems.

In this paper, we mainly focus on the \acf{MIA} against recommender systems.
To the best of our knowledge, ours is the first work to study debiasing learning for this task.
The most closely related work is~\citep{DBLP:conf/ccs/ZhangRWRCHZ21}.
However, the previous \ac{MIA} against recommender systems still face two challenging problems:
\begin{enumerate*}[label=(\roman*)]
    \item biased attack model training,
    \item inaccurate estimations of difference vectors.
\end{enumerate*}
In our proposed \ac{DL-MIA}, to mitigate the gap between target and shadow recommenders, the \ac{VAE} based encoder is employed to model recommender invariant and specific features.
In addition, to reduce the impacts of inaccurate estimations, the weight estimator is employed, and truth-level scores for difference vectors are calculated to facilitate the model update.


\section{Method}
\label{sec:Method}
We first formulate the \acf{MIA} against recommender systems.
Then, we give an overview of \ac{DL-MIA}.
Next, we explain \ac{DL-MIA}'s disentangled encoder and weight estimator.
Finally, the learning algorithm is presented.

\subsection{Problem formulation}
\label{subsec:Problem formulation}
Membership leakage in recommender systems happens when the adversary aims to determine whether a user's data is used to train the target recommender.
Formally, given a user's data $\mathbf{x}$, a trained target recommender $\mathcal{M}_{\mathit{target}}$, and external knowledge of the adversary $\Omega$, a membership inference attack model $\mathcal{A}$ can be defined as follows:
\begin{equation}
    \mathcal{A}:\mathbf{x}, \mathcal{M}_{\mathit{target}},\Omega \rightarrow \{0, 1\}, 
\end{equation}
where 0 means $\mathbf{x}$ is not a member of $\mathcal{M}_{\mathit{target}}$'s training dataset while 1 indicates $\mathbf{x}$ is a member.
The attack model $\mathcal{A}$ is essentially a binary classifier.

\header{Adversarial knowledge} 
In this attack, the adversary only has black-box access to the target recommender.
Specifically, only the recommendations to users, and users' historical behaviors (e.g., ratings or interaction sequences) are observational. 
In that case, as explained in~\citep{DBLP:conf/ccs/ZhangRWRCHZ21}, the adversary can infer the membership using the similarity between users' historical behaviors and recommended items from the target model.

\subsection{Model overview}
\label{subsec:Model overview}
\begin{figure*}[t]
    \centering
    \includegraphics[width=0.92\linewidth]{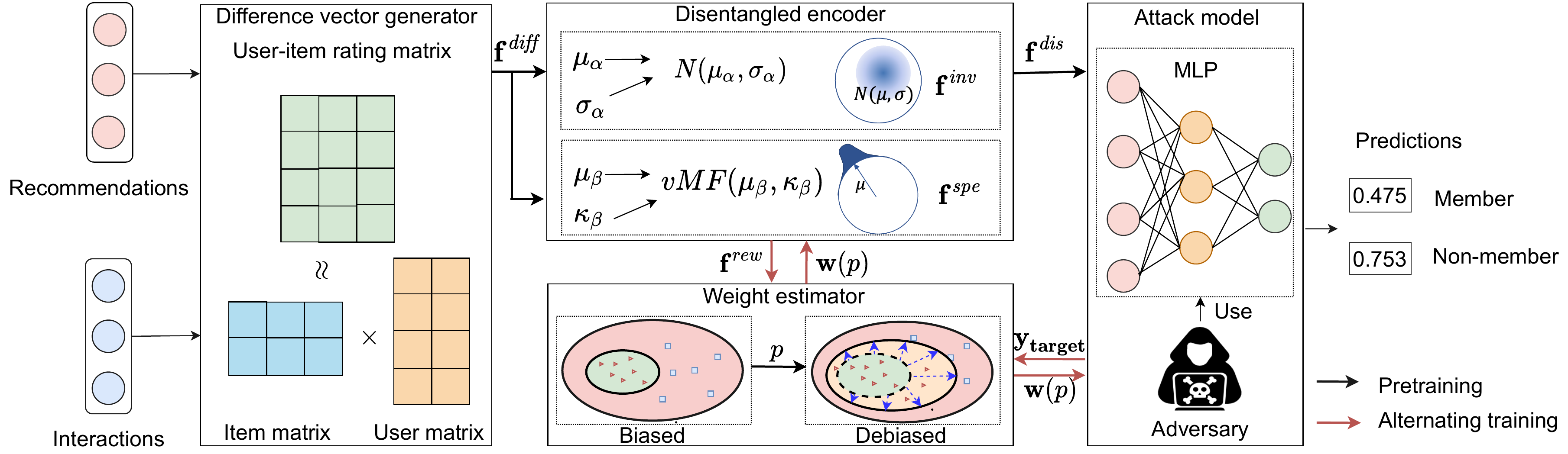}
    \caption{An overview of \ac{DL-MIA}. \ac{DL-MIA} has four main components: a difference vector generator, a disentangled encoder, a weight estimator, and an attack model.}
    \label{fig:An overview.}
  \vspace*{-.5\baselineskip}    
\end{figure*}

Figure~\ref{fig:An overview.} shows the four main components of \ac{DL-MIA}: 
\begin{enumerate*}[label=(\roman*)]
    \item a difference vector generator,
    \item a disentangled encoder,
    \item a weight estimator, and
    \item an attack model.
\end{enumerate*}
In this section, we give an overview of the \ac{DL-MIA} framework.

\subsubsection{Difference vector generator}
Following~\citet{DBLP:conf/ccs/ZhangRWRCHZ21}, to conduct \ac{MIA} against the target recommender, a shadow recommender $\mathcal{M}_{\mathit{shadow}}$ is established, and difference vectors between users' historical behaviors and recommended items are calculated.
To achieve this, we first factorize the user-item rating matrix to obtain item representations $\mathbf{M}^{\mathit{item}}$.
Then, a shadow recommender $\mathcal{M}_{\mathit{shadow}}$ is established and trained to simulate the target recommender.
Next, for the $i$-th user in $\mathcal{M}_{\mathit{shadow}}$, we project her/his interacted and recommended items into representations, denoted as $\mathbf{I}_{\mathit{shadow}, i}$ and $\mathbf{R}_{\mathit{shadow}, i}$, respectively. 
Finally, the difference vector for the $i$-th user is computed as:
\begin{equation}
\label{eq:difference vector generator}
    \mathbf{f}^{\mathit{diff}}_{\mathit{shadow}, i} = \overline{\mathbf{I}}_{\mathit{shadow},i} - \overline{\mathbf{R}}_{\mathit{shadow},i},
\end{equation}
where $\overline{\mathbf{I}}_{\mathit{shadow},i}$ and $\overline{\mathbf{R}}_{\mathit{shadow},i}$ are the averages of item vectors in $\mathbf{I}_{\mathit{shadow}, i}$ and $\mathbf{R}_{\mathit{shadow}, i}$, respectively.

\subsubsection{Disentangled encoder}
To mitigate the training data bias, the disentangled encoder aims to identify features invariant and specific to shadow and target recommenders.
Specifically, given generated difference vector $\mathbf{f}^{\mathit{diff}}$, a \ac{VAE} based encoder, composing two kinds of prior distributions, is employed to disentangle $\mathbf{f}^{\mathit{diff}}$ into the invariant feature $\mathbf{f}^{\mathit{inv}}$ and specific feature $\mathbf{f}^{\mathit{spe}}$.
And the disentangled difference vector $\mathbf{f}^{\mathit{dis}}$ is obtained by concatenating $\mathbf{f}^{\mathit{inv}}$ and $\mathbf{f}^{\mathit{spe}}$, i.e., $\mathbf{f}^{\mathit{dis}} = [\mathbf{f}^{\mathit{inv}};\mathbf{f}^{\mathit{spe}}]$.

\subsubsection{Weight estimator}
To further alleviate the influence of the estimation bias, the weight estimator assigns a truth-level score $p$ to each disentangled difference vector $\mathbf{f}^{\mathit{dis}}$. 
To learn $p$, the estimation constraint is constructed.
Moreover, to facilitate the model update and weight learning, an alternating training scheme is developed. 
In this way, the disentangled and reweighted difference vector $\mathbf{f}^{\mathit{rew}}$ is derived.

\subsubsection{Attack model}
For membership inference, a generic attack model $\mathcal{A}$ is essentially a binary classifier with the input of difference vectors.  
Following~\citet{DBLP:conf/ccs/ZhangRWRCHZ21}, we adopt a \ac{MLP} with 2 hidden layers for the attack model, i.e., $\mathcal{A}: \mathbf{y} = {\rm MLP}\left(\mathbf{f}^{\mathit{rew}}\right)$.
The output $\mathbf{y} = (y_{1}, y_{2})$ is a 2-dimensional vector indicating the probability of the user belonging to members ($y_{1}$) or non-members ($y_{2}$).
And the binary cross-entropy loss is used to train the attack model:
\begin{equation}
\label{eq:BCE}
    \mathcal{L}_{\mathit{BCE}} = -\sum_{i=1}^{N_{\mathit{shadow}}} \left(y_{i}^{*}\log y_{i, 1} + (1 -y_{i}^{*})\log y_{i, 2}\right),
\end{equation}
where $y_{i}^{*}$ is the ground truth label for $i$-th user, and $N_{\mathit{shadow}}$ is the size of training data generated by the shadow recommender.

Since we use the difference vector generator and attack model of the same architecture as the previous work~\citep{DBLP:conf/ccs/ZhangRWRCHZ21}, only the disentangled encoder and weight estimator are explained in detailed in the following sections.

\subsection{Disentangled encoder}
\label{subsec:Disentangled encoder}
Given the difference vector $\mathbf{f}^{\mathit{diff}}$ from the generator, the disentangled encoder aims to identify recommender invariant and specific features (i.e., $\mathbf{f}^{\mathit{inv}}$ and $\mathbf{f}^{\mathit{spe}}$).
To achieve this, inspired by~\citep{DBLP:conf/naacl/ChenTWG19}, we construct a \acf{VAE} using Gaussian and \ac{vMF} distributions to model recommender invariant and specific characteristics, respectively. 

Specifically, in the encoder, we assume a difference vector is generated by conditioning on two independent latent variables: the recommender invariant feature $\mathbf{f}^{\mathit{inv}}$ and the recommender specific feature $\mathbf{f}^{\mathit{spe}}$. 
Thus, the joint probability in our model is computed as follows:
\begin{equation}
    p_{\theta}(\mathbf{f}^{\mathit{diff}}, \mathbf{f}^{\mathit{inv}}, \mathbf{f}^{\mathit{spe}}) = p_{\theta}(\mathbf{f}^{\mathit{inv}}) p_{\theta}(\mathbf{f}^{\mathit{spe}}) p_{\theta}(\mathbf{f}^{\mathit{diff}} \vert \mathbf{f}^{\mathit{inv}}, \mathbf{f}^{\mathit{spe}}),
\end{equation}
where $p_{\theta}(\mathbf{f}^{\mathit{inv}})$ and  $p_{\theta}(\mathbf{f}^{\mathit{spe}})$ are the priors for $\mathbf{f}^{\mathit{inv}}$ and $\mathbf{f}^{\mathit{spe}}$, respectively.
And $p_{\theta}(\mathbf{f}^{\mathit{diff}} \vert \mathbf{f}^{\mathit{inv}}, \mathbf{f}^{\mathit{spe}})$ denotes the likelihood.
Following previous work~\cite{DBLP:conf/acl/ZhouN17,DBLP:conf/emnlp/ChenTLG18}, we assume a factored posterior probability $q_\phi(\mathbf{f}^{\mathit{inv}},\mathbf{f}^{\mathit{spe}}\vert  \mathbf{f}^{\mathit{diff}})=q_\phi(\mathbf{f}^{\mathit{inv}}\vert \mathbf{f}^{\mathit{diff}})q_\phi(\mathbf{f}^{\mathit{spe}}\vert \mathbf{f}^{\mathit{diff}})$. 
Therefore, learning of our encoder maximizes an evidence lower bound on marginal log-likelihood:
%
\begin{align}
    \mathcal{L}&_{\mathit{ELBO}} 
    \nonumber\\
    & \stackrel{\text{def}}{=\joinrel=} \mathop\mathbb{E}_{\substack{\mathbf{f}^{\mathit{inv}},
    \mathbf{f}^{\mathit{spe}}}}\Bigl[\log p_\theta\bigl(\mathbf{f}^{\mathit{diff}}\big\vert \mathbf{f}^{\mathit{spe}},\mathbf{f}^{\mathit{inv}}\bigr)
    -\log\frac{q_\phi(\mathbf{f}^{\mathit{spe}}\vert  \mathbf{f}^{\mathit{diff}})}{p_\theta(f^{\mathit{spe}})} 
    \nonumber \\
    & \qquad \qquad\,\, -\log\frac{q_\phi(\mathbf{f}^{\mathit{inv}}\vert  \mathbf{f}^{\mathit{diff}})}{p_\theta(\mathbf{f}^{\mathit{inv}})}\Bigr]
    \label{eq:ELBO}\\
     & =\!\! \mathop\mathbb{E}_{\substack{\mathbf{f}^{\mathit{inv}},
    \mathbf{f}^{\mathit{spe}}}}\! \bigl[\log p_\theta(\mathbf{f}^{\mathit{diff}}\vert  \mathbf{f}^{\mathit{spe}},\mathbf{f}^{\mathit{inv}})\bigr] 
    {-}\kld(q_\phi(\mathbf{f}^{\mathit{spe}}\vert  \mathbf{f}^{\mathit{diff}})\Vert p_\theta(\mathbf{f}^{\mathit{spe}}))  
    \nonumber \\
    &\quad -\kld\bigl(q_\phi(\mathbf{f}^{\mathit{inv}}\vert  \mathbf{f}^{\mathit{diff}})\Vert p_\theta(\mathbf{f}^{\mathit{inv}})\bigr),
    \nonumber
\end{align}
%
where $\mathbf{f}^{\mathit{inv}}\sim q_\phi(\mathbf{f}^{\mathit{inv}}\vert  \mathbf{f}^{\mathit{diff}})$ and $    \mathbf{f}^{\mathit{spe}}\sim q_\phi(\mathbf{f}^{\mathit{spe}}\vert  \mathbf{f}^{\mathit{diff}})$.
$q_\phi(\mathbf{f}^{\mathit{inv}}\vert  \mathbf{f}^{\mathit{diff}})$ and $q_\phi(\mathbf{f}^{\mathit{spe}}\vert  \mathbf{f}^{\mathit{diff}})$ are the posteriors.
$\kld(p \Vert q)$ denotes the KL divergence between the distribution $p$ and $q$.
In our disentangled encoder, two distribution families, i.e., the \ac{vMF} and Gaussian distributions, are used to define the posteriors. 
Further details on the parameterization are provided below.

\subsubsection{Gaussian Distribution.}
We assume that $q_\phi(\mathbf{f}^{\mathit{inv}}\vert  \mathbf{f}^{\mathit{diff}})$ follows a Gaussian distribution~\citep{DBLP:conf/aaai/QianFWC21} $\mathcal{N}(\mu_\beta(\mathbf{f}^{\mathit{diff}}),\text{diag}(\sigma_\beta(\mathbf{f}^{\mathit{diff}})))$, and that the prior $p_\theta(\mathbf{f}^{\mathit{inv}})$ follows the standard distribution $\mathcal{N}(0,I)$, where $I$ is an identity matrix.
In our encoder, we only consider a diagonal covariance matrix, and thus the KL divergence term $\kld(q_\phi(\mathbf{f}^{\mathit{inv}}\vert  \mathbf{f}^{\mathit{diff}})\Vert p_\theta(\mathbf{f}^{\mathit{inv}}))$ can also be obtained as follows:
\begin{equation}
    \frac{1}{2}\Bigl(-\sum_i\log\sigma_{\beta i} + \sum_i\sigma_{\beta i} + \sum_i{\mu_{\beta i} ^2} - d\Bigr).
\end{equation}

\subsubsection{vMF Distribution.} 
vMF can be recognized as a Gaussian distribution on a hypersphere with two parameters, $\mu$, and $\kappa$. 
$\mu\in\mathbb{R}^m$ is a normalized vector (i.e., $\Vert\mu\Vert_2=1$ ) and defines the mean direction. 
$\kappa\in\mathbb{R}_{\geq 0}$ denotes the concentration parameter analogous to the variance in a Gaussian distribution. 

In our encoder, we assume that $q_\phi(\mathbf{f}^{\mathit{spe}}\vert  \mathbf{f}^{\mathit{diff}})$ follows a \ac{vMF} distribution $\vmf(\mu_\alpha(\mathbf{f}^{\mathit{diff}}),\kappa_\alpha(\mathbf{f}^{\mathit{diff}}))$ and the prior $p_\theta(\mathbf{f}^{\mathit{spe}})$ follows the uniform distribution $\vmf(\cdot,0)$.
The $\kld(q_\phi(\mathbf{f}^{\mathit{spe}}\vert \mathbf{f}^{\mathit{diff}})\Vert p_\theta(\mathbf{f}^{\mathit{spe}}))$ term in $\mathcal{L}_{ELBO}$ can then be computed in closed form:
\begin{equation}
\begin{aligned}
    &\kappa_\alpha\frac{\mathcal{I}_{m/2}(\kappa_\alpha)}{\mathcal{I}_{m/2-1}(\kappa_\alpha)} + (m/2 - 1)\log\kappa_\alpha - 
    (m/2)\log(2\pi) \\
    &- \log\mathcal{I}_{m/2-1}(\kappa_\alpha)+
    \frac{m}{2}\log\pi+\log 2-\log\Gamma(m/2),
\end{aligned}
\end{equation}
where $\mathcal{I}_v$ is the modified Bessel function of the first kind at order $v$ and $\Gamma(\cdot)$ is the Gamma function. 
Following~\citet{DBLP:conf/uai/DavidsonFCKT18}, we use an acceptance-rejection scheme to sample from the vMF distribution.

\subsubsection{Reconstruction error}
We assume the conditional likelihood distribution $p_\theta(\mathbf{f}^{\mathit{diff}}\vert  \mathbf{f}^{\mathit{spe}},\mathbf{f}^{\mathit{inv}})$ follows $\mathcal{N}(f([\mathbf{f}^{\mathit{inv}};\mathbf{f}^{\mathit{spe}}]), \mathbf{I})$, where a \ac{MLP} with 3 hidden layers is adopted for $f(\cdot)$. 
Thus, the reconstruction error (the first term) in $\mathcal{L}_{ELBO}$ can be rewritten as:
\begin{equation}
    \mathop\mathbb{E}_{\substack{\mathbf{f}^{\mathit{inv}},
    \mathbf{f}^{\mathit{spe}}}}\left[-\frac{1}{2}\left\lVert f([\mathbf{f}^{\mathit{inv}};\mathbf{f}^{\mathit{spe}}]) - \mathbf{f}^{\mathit{diff}}\right\rVert ^{2}\right].
\end{equation}
During training, we use a linear layer to produce $\mu_{\beta}$, $\delta_{\beta}$, $\mu_\alpha$, and $\kappa_{\alpha}$.
The difference vectors from both shadow and target recommenders are disentangled by the encoder.
Note that membership labels in the target recommender are not exposed to \ac{DL-MIA}.
Through the encoder, the disentangled difference vector, i.e., $\mathbf{f}^{\mathit{dis}} = [\mathbf{f}^{\mathit{inv}};\mathbf{f}^{\mathit{spe}}]$, is obtained to mitigate the gap between shadow and target recommender.

\subsection{Weight estimator}
\label{subsec:Weight estimator}
Given $\mathbf{f}^{\mathit{dis}}$ from the disentangled encoder, the weight estimator aims to alleviate the estimation bias of difference vectors.
Specifically, we introduce the truth-level score to indicate the estimation accuracy.
Then, we establish the estimation constraint, and assign a truth-level score for each difference vector.
Moreover, to update model parameters and learn scores simultaneously, an alternating training strategy among the disentangled encoder, weight estimator, and attack model is adopted.

\subsubsection{Truth-level score}
As mentioned in Sec.~\ref{subsec:Problem formulation}, the hidden states in the target recommender, including item representations, are not observational to the adversary.
As a result, difference vectors for recommenders may be computed inaccurately by \ac{MF} in the generator.
In the estimator, we write $\mathbf{f'}$ for the ground truth difference vector, and define the truth-level score $p$ for $\mathbf{f}^{\mathit{dis}}$ as follows:
\begin{equation}
\label{eq:truth-level score}
    p = \frac{\delta \left(\mathcal{A}\left(\mathbf{f'}\right), y^{*}\right)}{\delta (\mathcal{A}(\mathbf{f}^{\mathit{dis}}), y^{*})},
\end{equation}
where $\mathcal{A}(\cdot)$ denotes the attack model, and $y^{*}$ is the membership label. $\delta(\cdot)$ is the error measure, for which we adopt the binary cross-entropy loss.

\subsubsection{Alternating training}
In the estimator, the truth-level score serves as the weighting parameter for the estimation bias.
After debiasing by the truth-level score, the biased estimation should be equal to the unbiased estimation.
Motivated by this, we can rewrite Eq.~\ref{eq:truth-level score} as follows:
\begin{equation}
    p \cdot\delta \left(\mathcal{A}\left(\mathbf{f}^{\mathit{dis}}\right), y^{*}\right) = \delta \left(\mathcal{A}\left(\mathbf{f'}\right), y^{*}\right).
\end{equation}
On this basis, to compute the truth-level score $p$, the estimation constraint is established:
\begin{equation}
\label{eq:estimation constraint}
\begin{split}
   &\mathcal{L}_{\mathit{estimate}}  = {}\\
   &\sum_{j}{\lambda_{j} {\cdot} \!\sum_{i=1}^{N_{j}}{\! \left\lVert p_{i,j} {\cdot} \delta \left(\mathcal{A}\left(\mathbf{f}_{i,j}^{\mathit{dis}}\right), y_{i,j}^{*}\right) - \delta \left(\mathcal{A}\left(\!\mathbf{f'}_{i,j}\right), y_{i.j}^{*}\right)\right\rVert^{2}}},
   \end{split}
\end{equation}
where $j \in \{shadow, target\}$, and $\lambda_{j}$ is the weight for the shadow or target recommender.
Here, we set $\lambda_{j} = \frac{1}{N_{j}}$, where
$N_{\mathit{shadow}}$ and $N_{\mathit{target}}$ are the size of training dataset generated by shadow recommender, and the test dataset for the target recommender, respectively.

However, membership labels $\mathbf{y}^{*}_{\mathit{target}}$ of the target recommender and the ground truth difference vector $\mathbf{f}'$ cannot be obtained directly.
To address this issue and facilitate model update, we develop an alternating training strategy among the disentangled encoder, attack model, and weight estimator.
Specifically, the re-weighted loss for the disentangled encoder and attack model is defined as follows:
\begin{align}
    \mathcal{L}_{\mathit{reweight}} &= \mathcal{L}'_{BCE} + \mathcal{L}'_{ELBO},
    \label{eq:reweight loss} \\
    \mathcal{L}'_{BCE} &= -\!\!\sum_{i=1}^{N_{\mathit{shadow}}}\!\! \mathbf{w}_{\mathit{shadow}, i}(\mathbf{p}) \cdot \left(y_{i}^{*}\log y_{i, 1} + (1 -y_{i}^{*})\log y_{i, 2}\right),
    \nonumber\\
    \mathcal{L}'_{\mathit{ELBO}} &= -\sum_{j}{\sum_{i}^{N_{j}}{\mathbf{w}_{j,i}(\mathbf{p})\cdot \mathcal{L}_{{ELBO,j,i}}}},\quad j\in \{\mathit{shadow}, \mathit{target}\},
    \nonumber
\end{align}
%
where $\mathbf{w}_{\mathit{shadow}}(\mathbf{p})$ and $\mathbf{w}_{\mathit{target}}(\mathbf{p})$ are the data sample weights for shadow and target recommenders, obtained by applying a linear layer on the current truth-level scores $\mathbf{p}$.
We compute the re-weighted and disentangled difference vector $\mathbf{f}^{\mathit{rew}} = \left[\mathbf{f}'^{\mathit{inv}};\mathbf{f}'^{\mathit{spe}}\right]$ by minimizing $\mathcal{L}_{\mathit{reweight}}$, where $\mathbf{f}'^{\mathit{inv}}$ and $\mathbf{f}'^{\mathit{spe}}$ are the re-weighted invariant and specific vectors.
Meanwhile, the trained attack model is able to predict membership labels for the target recommender, i.e., $y_{\mathit{target}}$.
Next, we approximate $\mathbf{f}'$ and $y^{*}_{\mathit{target}}$ by $\mathbf{f}^{\mathit{rew}}$ and $y_{\mathit{target}}$, and minimize $\mathcal{L}_{\mathit{estimate}}$ to refine the current truth-level scores $\mathbf{p}$.
In this way, $\mathcal{L}_{\mathit{reweight}}$ and $\mathcal{L}_{\mathit{estimate}}$ are optimized in an alternating fashion.

\subsection{Learning algorithm}
\label{subsec:Learning algorithm}
The training process of \ac{DL-MIA} contains two stages:
\begin{enumerate*}[label=(\roman*)]
    \item \textbf{Pretraining}. We first pretrain the disentangled encoder and attack model jointly by optimizing $\mathcal{L}_{\mathit{BCE}}$ (Eq.~\ref{eq:BCE}) and $\mathcal{L}_{\mathit{ELBO}}$ (Eq.~\ref{eq:ELBO}).
    In this stage, the disentangled difference vector $\mathbf{f}^{\mathit{dis}}$ is computed, and inputted into the attack model for learning.
    \item \textbf{Alternating training}. After obtaining the disentangled difference vectors, we adopt the alternating training strategy to reduce the estimation bias. 
    Specifically, the re-weighted loss $\mathcal{L}_{\mathit{reweight}}$ and the estimation constraint $\mathcal{L}_{\mathit{estimate}}$ are minimized iteratively.
    In such manner, the re-weighted difference vector $\mathbf{f}^{\mathit{rew}}$ is derived, and then used by the attack model to conduct membership inference on the target recommender.
\end{enumerate*}
Sec.~\ref{subsec:learning algorithm and training} gives the detailed training algorithm of \ac{DL-MIA}.


\section{Experiments}
\label{sec:Experiments}

\textbf{Research questions.}
We aim to answer the following research questions:
\begin{enumerate*}[label=(RQ\arabic*),leftmargin=*]
\item Does \ac{DL-MIA} outperform the state-of-the-art attack methods? 
Is \ac{DL-MIA} able to generalize to the sequential recommendation? (Sec.~\ref{subsec:MIA against RS} and ~\ref{subsec:MIA against SR}) 
\item How does the disentangled encoder and weight estimator contribute to the performance? (Sec.~\ref{subsec:ablation study})
\item What is the influence of the difference vector generator and defense mechanism? (Sec.~\ref{subsec:Influence of difference vector generator} and~\ref{subsec:Influence of defense mechanism})
\item Is \ac{DL-MIA} able to identify features invariant and specific to shadow and target recommenders, and alleviate the estimation bias? (Sec.~\ref{subsec:Case studies})
\end{enumerate*}

\header{Datasets}
Following~\citet{DBLP:conf/ccs/ZhangRWRCHZ21}, we evaluate the attack performance on two real-world datasets, MovieLens~\citep{DBLP:journals/tiis/HarperK16} and Amazon~\citep{DBLP:conf/sigir/McAuleyTSH15}.
MovieLens is a widely used benchmark dataset for evaluating collaborative filtering algorithms. We use the version (MovieLens-1M) that includes 1 million user ratings for both general and sequential recommenders.
Amazon is a series of datasets, consisting of large corpora of product reviews crawled from Amazon.com. 
Top-level product categories on Amazon are treated as separate datasets. 
Similar to~\citet{DBLP:conf/ccs/ZhangRWRCHZ21}, we consider ``Digital Music'' for general recommenders, while ``Beauty'' is used for sequential recommenders, since the number of the interacting users per item in ``Digital Music'' is extremely low (less than 2). 
Table~\ref{tab:Statistics of processed datasets for recommender systems.} summarizes the statistics of datasets.

Similar to~\citep{DBLP:conf/ccs/ZhangRWRCHZ21}, we further divide each dataset into three disjoint subsets. i.e., a shadow dataset, a target dataset, and a dataset for difference vector generation. 
We filter the target and shadow datasets to make sure the dataset for difference vector generation contains all the items.
Then, target and shadow datasets are both randomly separated into two disjoint parts for members and non-members, respectively.
For general recommenders, we remove users with less than 20 interactions.
For sequential recommenders, we filter out users and items with less than 5 interaction records.

\begin{table}[t]
  \centering
  \caption{Statistics of datasets. \#Users, \#Items, and \#Interactions denote the number of users, items, and user-item interactions, respectively.}
  \label{tab:Statistics of processed datasets for recommender systems.}
  \begin{tabular}{l rrr}
  \toprule
    Dataset & \#Users & \#Items & \#Interactions  \\
  \midrule
    MovieLens-1M & 6,040 & 3,706 & 1,000,209 \\
    Amazon Digital Music & 840,372 & 456,992 & 1,584,082  \\
    Amazon Beauty & 1,210,271 & 249,274 & 2,023,070  \\
  \bottomrule
\end{tabular}
\vspace*{-.5\baselineskip}
\end{table}

\header{Recommender systems}
Following~\citep{DBLP:conf/ccs/ZhangRWRCHZ21}, we evaluate membership inference attacks against three general recommenders:
\begin{enumerate*}[label=(\roman*)]
    \item \acf{ItemBase}~\citep{DBLP:conf/www/SarwarKKR01},
    \item \acf{LFM} \citep{DBLP:conf/sac/PolatD05}, and 
    \item \acf{NCF}~\citep{he2017neural}.
\end{enumerate*}
To investigate the generality of our proposed model, we also implement attack models on three sequential recommendation methods in our experiments, including GRU4Rec~\citep{DBLP:journals/corr/HidasiKBT15}, Caser~\citep{DBLP:conf/wsdm/TangW18},  and BERT4Rec~\citep{DBLP:conf/cikm/SunLWPLOJ19}.

\header{Experimental settings}
Table~\ref{tab:Notations for different settings.} shows the notation for our experimental settings. 
Note that not all possible settings are listed due to space limitations.
In the experiments, there are two kinds of combinations (i.e., 2-letter and 4-letter combinations) for experimental settings. 
For the 2-letter combinations, the first letter indicates the dataset, and the second letter denotes the recommendation algorithm. 
For example, for general recommenders, ``AI'' denotes that the recommender is implemented by ItemBase and trained on Amazon Digital Music.
For the 4-letter combinations, the first two letters represent the dataset and algorithm used by the shadow recommender, and the last two letters denote the dataset and algorithm used by the target recommender. 
For instance, for sequential recommenders, ``ABMC'' means the adversary establishes a shadow recommender using BERT4REC on Amazon Beauty to attack a target recommender using Caser on MovieLens-1M.

\header{Baseline}
\label{subsec:baselines}
We compare the proposed \ac{DL-MIA} with the biased baseline (Biased)~\citep{DBLP:conf/ccs/ZhangRWRCHZ21}, which is the first work studying the membership inference attack against recommender systems.
Previous \ac{MIA} methods~\citep{DBLP:conf/sp/ShokriSSS17,DBLP:conf/ndss/Salem0HBF019,DBLP:conf/csfw/YeomGFJ18, DBLP:conf/sp/NasrSH19, DBLP:conf/icml/Choquette-ChooT21,DBLP:conf/ccs/LiZ21} are not considered in our experiments, since they cannot be directly applied to recommender systems.

\header{Evaluation metric}
We adopt the area under the ROC curve (AUC) as the evaluation metric. 
AUC signifies the probability that the positive sample’s score is higher than the negative sample’s score, illustrating the classification model’s ability to rank samples.
For example, if the attack model infers the membership with random guessing, the AUC is close to 0.5.

\header{Implementation details}
\label{subsec:implementation details}
For the attack model, we build a \ac{MLP} with 2 hidden layers. 
And the first layer has 32 units and the second layer has 8 units. 
We employ the ReLU activation function, and use the softmax function as the output layer.
For optimizers, we employ Adam with a learning rate of 0.001 for the disentangled encoder and SGD with a learning rate of 0.01 and a momentum of 0.7 for the attack model. 
During training, we first pretrain the attack model and the disentangled encoder jointly for 200 epochs. 
Then, truth-level scores and model parameters are alternatively updated for every 10 epochs. 
The whole alternative training is conducted for 100 epochs.
Following~\citep{DBLP:conf/ccs/ZhangRWRCHZ21}, we consider the top 100 recommendations for members, and recommend the most popular items to non-members.  
Table~\ref{tab:Parameter settings of different recommender systems} lists detailed parameter settings.


\section{Experimental Results}
\label{sec:Results}
For RQ1, we evaluate the attack performance of our proposed \ac{DL-MIA} over general and sequential recommender systems.

\subsection{Attack performance over general recommenders (RQ1)}
\begin{figure}
  \centering
  \subfigure[General recommenders]{
    \includegraphics[clip, trim=0mm 1mm 0mm 0mm, width=\linewidth]{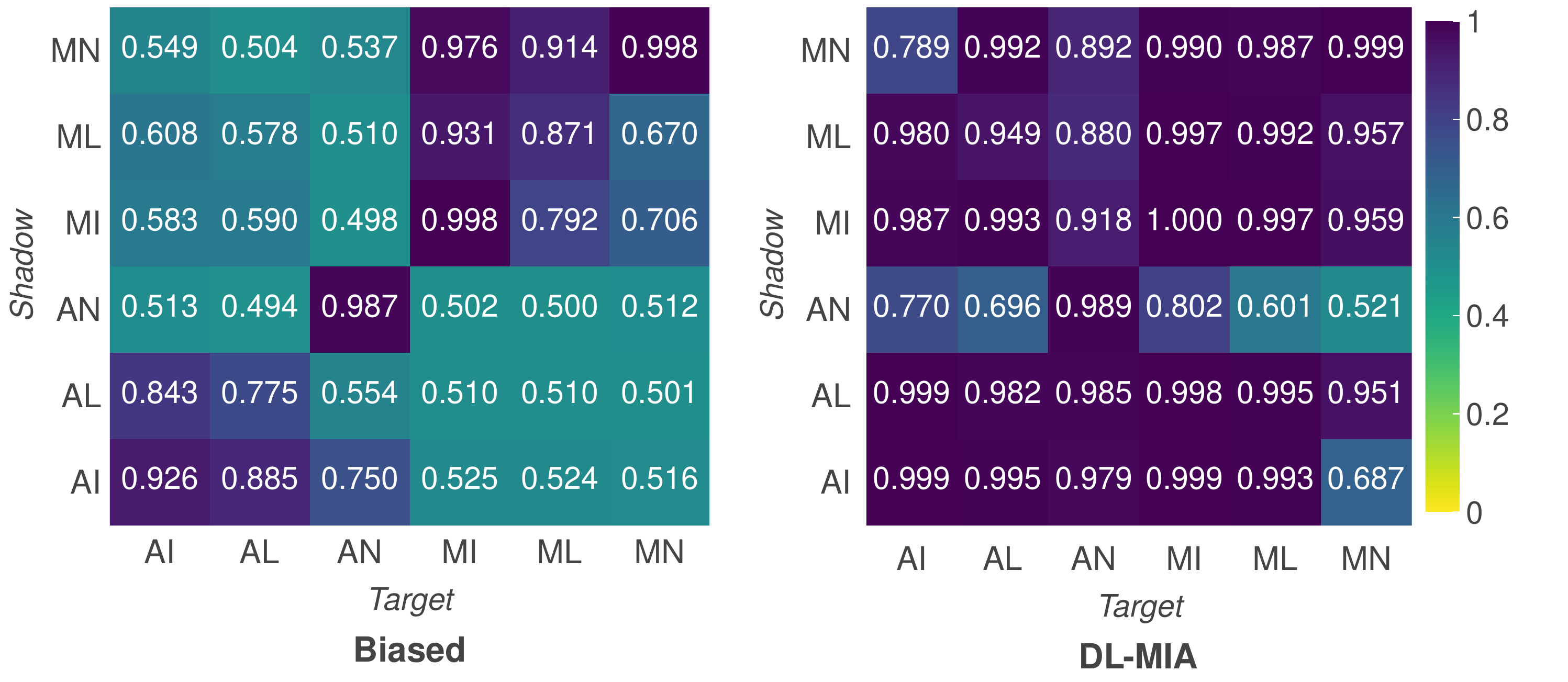}
    \label{fig:heatmap_RS}}
    \vspace*{-2mm}
    \\
  \subfigure[Sequential recommenders]{
    \label{fig:heatmap_SR}
    \includegraphics[width=\linewidth]{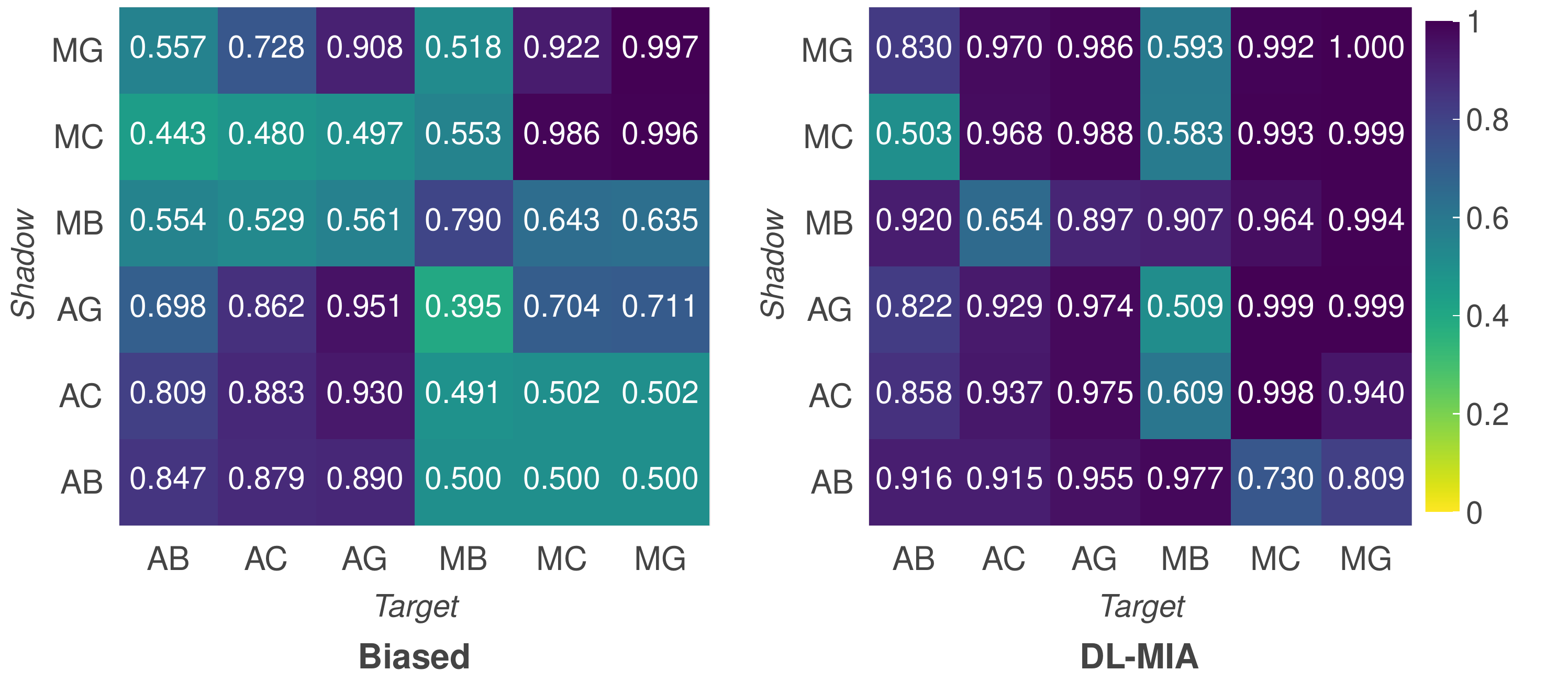}
    \label{fig:heatmap_SR}}
    \vspace*{-2mm}    
  \caption{Attack performance (AUC) over general recommenders (a) and sequential recommenders (b).}
  \label{fig:heatmap}
  \vspace*{-.5\baselineskip}
\end{figure}

Figure~\ref{fig:heatmap_RS} shows the experimental outcomes for the attack performance over general recommender systems.
Based on the experimental results, we have the following observations:
\begin{enumerate*}[label=(\roman*)]
    \item Membership inference attack against general recommender systems is challenging, and for the biased baseline, AUC scores are less than 0.7 in most settings (more than 60\%).
    In contrast, our proposed \ac{DL-MIA} is able to effectively infer the membership of the target recommenders, and AUC scores are over 0.8 for more than 80\% experimental settings.
    \item The proposed \ac{DL-MIA} consistently outperforms the biased baseline in all the settings.
    For example, for the ``MLAI'' setting, the AUC score of \ac{DL-MIA} is 0.980, while that of the biased baseline attack is 0.608.
    That is, identifying features invariant and specific to recommenders, and computing the truth-level scores for difference vectors substantially enhance the attack performance.
    \item Similar to the conclusions mentioned in~\citep{DBLP:conf/ccs/ZhangRWRCHZ21}, with knowledge of the algorithm and dataset distribution used by the target recommender, the adversary is capable of conducting a strong attack, and AUC scores of \ac{DL-MIA} at the back-diagonal in Figure~\ref{fig:heatmap_RS} are the highest in most cases.   
\end{enumerate*}
In summary, the proposed \ac{DL-MIA} can effectively infer the membership for the target recommender. 
Identifying recommender invariant and specific features, as well as considering estimation accuracy of difference vectors, are beneficial for the membership inference attack.

\label{subsec:MIA against RS}

\subsection{Attack performance over sequential recommenders (RQ1)}
\label{subsec:MIA against SR}
To investigate the generality of \ac{DL-MIA}, we report the performance of membership inference attacks against sequential recommenders.
Based on the results in Figure~\ref{fig:heatmap_SR}, we arrive at the following insights:
\begin{enumerate*}[label=(\roman*)]
    \item Even with ordered user historical behaviors, \ac{DL-MIA} can still accurately calculate the difference vectors, and infer  membership. 
    The AUC scores of \ac{DL-MIA} are over 0.8 for more than 80\% settings.
    \item \ac{DL-MIA} surpasses the biased baseline, and achieves better attack performance over both general and sequential recommenders, demonstrating the effectiveness and strong generalizability of our proposed framework.
\end{enumerate*}
In summary, the \ac{DL-MIA} framework cannot only effectively conduct membership inference attacks against general recommenders, but also attains the best AUC scores over sequential recommenders.


\section{Analysis}
\label{sec:Analysis}
In this section, we take a closer look at \ac{DL-MIA} to analyze its performance.
We examine how the disentangled encoder, and the weight estimator contribute to the performance.
The influence of the difference vector generator and defense mechanism is also investigated.
In addition, we conduct case studies to study whether \ac{DL-MIA} is able to recognize recommender invariant and specific features, and mitigate the estimation bias.

\subsection{Ablation studies (RQ2)}
We conduct ablation studies over both general and sequential recommenders.
The results are shown in Table~\ref{tab:Ablation studies over recommenders.}.
AUC scores are adopted here to evaluate the attack performance.
When only employing the difference vector generator and attack model, our framework is reduced to the biased baseline.
In that case, AUC scores over all the settings suffer a dramatic drop.
In the ``-Reweight'' setting, the disentangled encoder and attack model are trained jointly whereas the alternative training is removed.
Compared to the biased baseline, identifying features invariant and specific to shadow and target recommenders considerably alleviates the training data bias, and AUC scores are consistently improved over both general and sequential recommenders.
Meanwhile, \ac{DL-MIA} further enhances the attack performance by reducing the estimation bias of difference vectors.
In a nutshell, both the disentangled encoder and weight estimator contribute to the improvements in attack performance.

In Tabe~\ref{tab:Ablation studies over recommenders.}, we also consider two model variants, \ac{DL-MIA} ($\beta$-VAE) and \ac{DL-MIA} (FactorVAE), with two widely-used \ac{VAE} models $\beta$-VAE~\citep{DBLP:conf/iclr/HigginsMPBGBML17} and FactorVAE~\citep{DBLP:conf/icml/KimM18}, respectively.
Based on the results in Table~\ref{tab:Ablation studies over recommenders.}, we observe that \ac{DL-MIA} ($\beta$-VAE), and \ac{DL-MIA} (FactorVAE) both achieve a similar attack performance as \ac{DL-MIA}.
That is, even with a different \ac{VAE} based encoder, our proposed framework is still able to perform effective membership inference.

\label{subsec:ablation study}
\begin{table}
	\centering
	\caption{Ablation studies over general and sequential recommenders. \ldots ($\beta$-VAE) and \ldots (FactorVAE) are two model-variants of DL-MIA with widely-used VAE models.}
 	\setlength{\tabcolsep}{0.5mm}
	\begin{tabular}{l cccc cccc}
	\toprule
	\multirow{2}{*}{Model} &\multicolumn{4}{c}{General} &\multicolumn{4}{c}{Sequential}\\
	\cmidrule(lr){2-5} \cmidrule(lr){6-9}
	 & \rotatebox[origin=c]{60}{AIMI} & 
	 \rotatebox[origin=c]{60}{ALML}  & 
	 \rotatebox[origin=c]{60}{AIML} & 
	 \rotatebox[origin=c]{60}{ALMI} 	& 
	 \rotatebox[origin=c]{60}{AGMG} & 
	 \rotatebox[origin=c]{60}{MGAG}  & 
	 \rotatebox[origin=c]{60}{MGMC} & 
	 \rotatebox[origin=c]{60}{AGMC}
	  \\ \midrule
    DL-MIA & \textbf{0.999} & \textbf{0.995} & 0.993 & 0.998 & \textbf{0.999} & \textbf{0.986}  & 0.992 & \textbf{0.999} \\ 
	\midrule
	-Reweight & 0.553 & 0.518 & 0.531 & 0.526 & 0.723 & 0.915 & 0.936 & 0.707\\
	Biased~\citep{DBLP:conf/ccs/ZhangRWRCHZ21}     & 0.525 & 0.510  & 0.524 & 0.510 & 0.711 & 0.908  & 0.922 & 0.704 \\ 
	\midrule
	\ldots ($\beta$-VAE)        & 0.920 & 0.993   & 0.995 & 0.996 & 0.999 & 0.983 &0.993 & 0.996\\
	\ldots (FactorVAE)  & 0.999 & 0.987 &\textbf{0.999} &\textbf{0.999}   & 0.999 & 0.982  &\textbf{0.994}       & 0.999\\
	\bottomrule
	\end{tabular}
	\label{tab:Ablation studies over recommenders.}
  \vspace*{-.5\baselineskip}	
\end{table}

\subsection{Influence of difference vector generator (RQ3)}
\label{subsec:Influence of difference vector generator}
Table~\ref{tab:Influence of the difference vector generator over general recommender systems.} shows the attack performance (AUC) with two kinds of difference vector generators, ``MF'' and ``BERT''.
``MF'' generates the difference vectors by factorizing user-item matrices (explained in Sec.~\ref{subsec:Model overview}). ``BERT'' employs the tiny-sized BERT~\citep{DBLP:conf/naacl/DevlinCLT19} to embed item descriptions, and takes the [CLS] vectors as item representations.
Since some item descriptions are missing, we do not consider experimental settings using the Amazon Digital Music dataset.
Based on the results in Table~\ref{tab:Influence of the difference vector generator over general recommender systems.}, we find that \ac{DL-MIA} performs better than the biased baseline over both generators, indicating the effectiveness of our framework.

\begin{table}
	\centering
	\caption{Influence of the difference vector generator over general and sequential recommenders. ``Gen.'' is short for ``Generator.''}
	\setlength{\tabcolsep}{0.5mm}
	\begin{tabular}{l l ccccc cccc}
		\toprule
	\multirow{4}{*}{\rotatebox[origin=c]{90}{Gen.} } &  &\multicolumn{4}{c}{General} &\multicolumn{4}{c}{Sequential}\\
	\cmidrule(lr){3-6} \cmidrule(lr){7-10}
		 & Model & 
		\rotatebox[origin=c]{60}{MIMI} & 
		\rotatebox[origin=c]{60}{MIMN} & 
		\rotatebox[origin=c]{60}{MLMI} & 
		\rotatebox[origin=c]{60}{MNML} & 
		\rotatebox[origin=c]{60}{AGMG} & 
		\rotatebox[origin=c]{60}{MGAG} & 
		\rotatebox[origin=c]{60}{MGMC} & 
		\rotatebox[origin=c]{60}{AGMC} \\
		\midrule
		\multirow{2}{*}{\rotatebox[origin=c]{90}{\ac{MF}}}
		& Biased~\citep{DBLP:conf/ccs/ZhangRWRCHZ21} & 0.998 & 0.706 & 0.931 & 0.914  & 0.711 & 0.908  & 0.922 & 0.704 \\
		& DL-MIA & \textbf{1.000}  & \textbf{0.959} & \textbf{0.997} & \textbf{0.987}  & \textbf{0.999} & \textbf{0.986}  & \textbf{0.992} & \textbf{0.999} \\ 
		\midrule
		\multirow{2}{*}{\rotatebox[origin=c]{90}{BERT}}
		& Biased~\citep{DBLP:conf/ccs/ZhangRWRCHZ21} 
		         & 0.979 & 0.575 & 0.505 & 0.706   & 0.517 & 0.584  & 0.922 & 0.621\\
		& DL-MIA & \textbf{0.980} & \textbf{0.830} & \textbf{0.971} & \textbf{0.770} &\textbf{0.873}  &\textbf{0.863}   & \textbf{0.979}  &\textbf{0.870} \\ 
		\bottomrule
	\end{tabular}
	\label{tab:Influence of the difference vector generator over general recommender systems.}
\end{table}

\subsection{Influence of defense mechanism (RQ3)}
\label{subsec:Influence of defense mechanism}
\begin{figure}
  \centering
    \includegraphics[width=\linewidth]{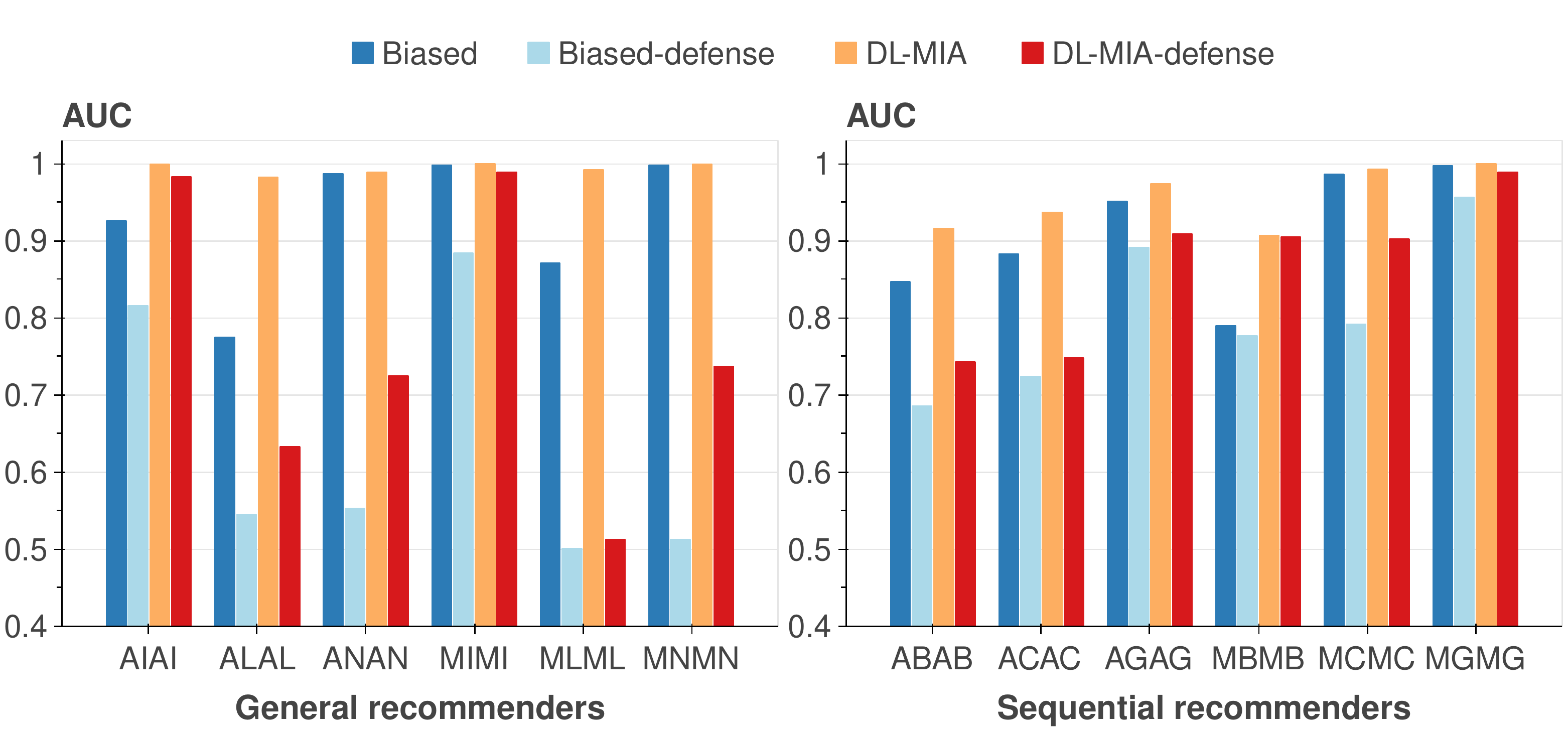}
    \caption{Influence of the defense mechanism. ``Biased-defense'' and ``DL-MIA-defense'' denote Biased~\citep{DBLP:conf/ccs/ZhangRWRCHZ21} and \ac{DL-MIA} with the defense mechanism, respectively.}
  \label{fig:Influence of the defense mechanism.}
\end{figure}

Following~\citet{DBLP:conf/ccs/ZhangRWRCHZ21}, we investigate the influence of defense mechanism, and apply the countermeasure named \emph{Popularity Randomization} to the attack frameworks. 
Figure~\ref{fig:Influence of the defense mechanism.} shows the attack performance (AUC) before and after deploying the defense mechanism.
With the defense mechanism, the attack performance for both the biased baseline and \ac{DL-MIA} consistently decreases over all the settings.
Meanwhile, compared to the biased baseline, our proposed \ac{DL-MIA} achieves higher AUC scores, and shows a stronger robustness to the countermeasure.

\subsection{Case studies (RQ4)}
\label{subsec:Case studies}
\begin{figure}
  \centering
  \subfigure[Invariant feature $\mathbf{f}^{\mathit{inv}}$]{
    \label{fig:invariant.}
    \includegraphics[clip, trim=0mm 5mm 0mm 0mm 0mm, width=0.485\linewidth]{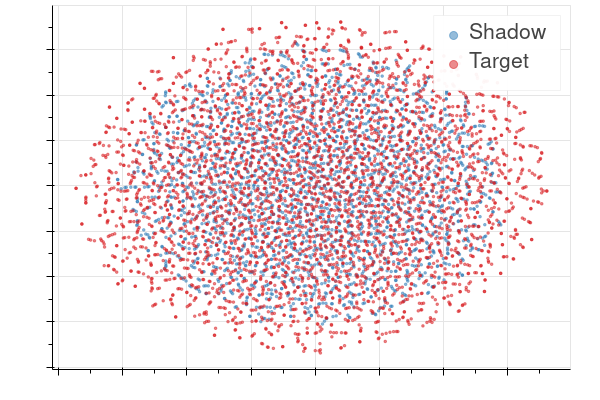}}
  \subfigure[Specific feature $\mathbf{f}^{\mathit{spe}}$]{
    \label{fig:specific.}
    \includegraphics[clip, trim=0mm 5mm 0mm 0mm 0mm, width=0.485\linewidth]{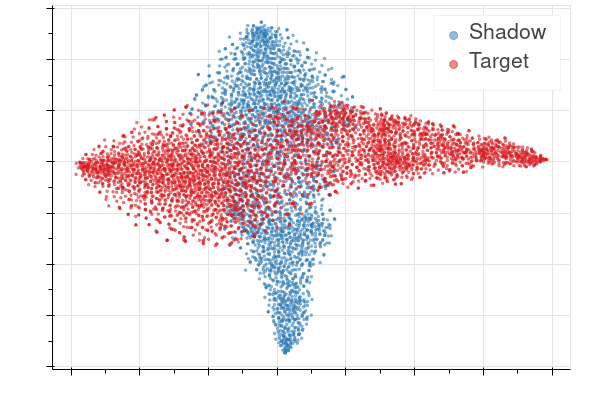}}
  \subfigure[Biased difference vector]{
    \label{fig:difference vectors before debiasing.}
    \includegraphics[clip, trim=0mm 5mm 0mm 0mm 0mm, width=0.485\linewidth]{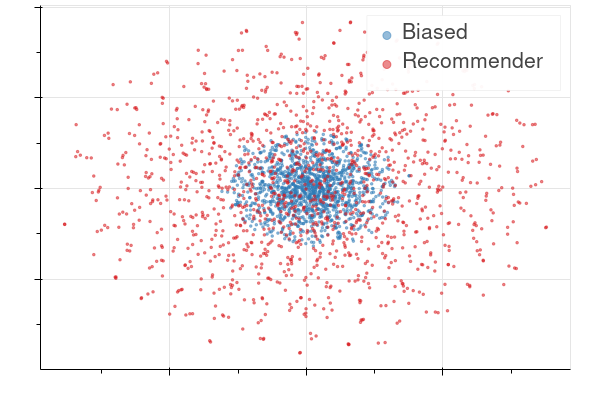}}
   \subfigure[Debiased difference vector]{
    \label{fig:difference vectors after debiasing.}
    \includegraphics[clip, trim=0mm 5mm 0mm 0mm 0mm, width=0.485\linewidth]{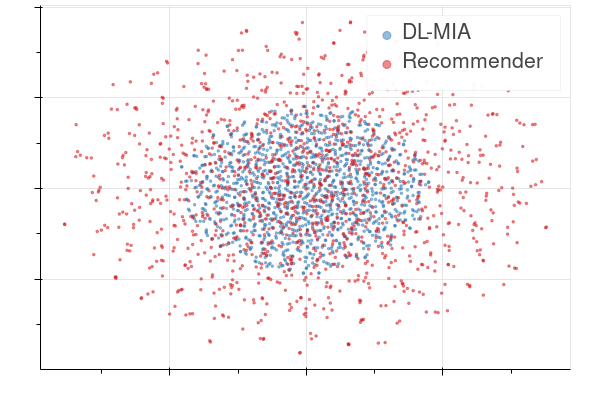}}
  \caption{Visualization results of the ``ANML'' setting for the general recommender.}
  \label{fig:case studies}
\end{figure}

Figure~\ref{fig:case studies} shows visualization results of the ``ANML'' setting for the general recommender system by t-SNE~\citep{JMLR:v9:vandermaaten08a}.
Points in Figure~\ref{fig:invariant.} and Figure~\ref{fig:specific.} stand for the recommender invariant and specific features, respectively.
We can see that invariant features from the shadow recommender (red) and target recommender (blue) are distributed similarly, whereas specific features are scattered divergently.
That is, by employing the disentangled encoder, \ac{DL-MIA} is able to mitigate the gap between recommenders.

In addition, Figure~\ref{fig:difference vectors before debiasing.} and Figure~\ref{fig:difference vectors after debiasing.} demonstrate the visualization results of difference vectors before and after debiasing, respectively.
Based on the results, we conclude that \ac{DL-MIA} effectively reduce the gap between difference vectors generated by the attack model (red) and recommender (blue), and alleviate the influence of the estimation bias.

\section{Conclusion and future work}
\label{sec:Conclusion}
In this paper, we investigate the membership inference attack against recommender systems.
Previously published methods faces two challenging problems:
\begin{enumerate*}[label=(\roman*)]
    \item the biased attack model training caused by the gap between target and shadow recommenders,
    \item and inaccurate estimation of difference vectors since hidden states in recommenders are inaccessible.
\end{enumerate*}
To handle these problems, we propose a novel framework named \ac{DL-MIA}.
To mitigate the gap between target and shadow recommenders, the \ac{VAE} based encoder is devised to identify recommender invariant and specific features.
And to alleviate the estimation bias, the weight estimator is constructed, and truth-level scores for difference vectors are computed to facilitate the model update.
We evaluate \ac{DL-MIA} against both general recommenders and sequential recommenders on two real-world datasets.
Experimental results demonstrate that the \ac{DL-MIA} framework is able to effectively alleviate training and estimation biases, and shows a strong generality.

In future work, we intend to incorporate more kinds of disentangled methods and explore other types of biases in the membership inference attack against recommender systems.

\begin{acks}
This work was supported by the National Key R\&D Program of China with grant No. 2020YFB1406704, the Natural Science Foundation of China (61902219, 61972234, 62072279, 62102234), the Natural Science Foundation of Shandong Province (ZR2021QF129), the Key Scientific and Technological Innovation Program of Shandong Province (2019JZZY010129), Shandong University multidisciplinary research and innovation team of young scholars (No. 2020QNQT017), Meituan, the Hybrid Intelligence Center, a 10-year program funded by the Dutch Ministry of Education, Culture and Science through the Netherlands Organisation for Scientific Research, \url{https://hybrid-intelligence-centre.nl}. 
All content represents the opinion of the authors, which is not necessarily shared or endorsed by their respective employers and/or sponsors.
\end{acks}

\clearpage
\bibliographystyle{ACM-Reference-Format}
\balance
\bibliography{main}


\begin{thebibliography}{55}


\ifx \showCODEN    \undefined \def \showCODEN     #1{\unskip}     \fi
\ifx \showDOI      \undefined \def \showDOI       #1{#1}\fi
\ifx \showISBNx    \undefined \def \showISBNx     #1{\unskip}     \fi
\ifx \showISBNxiii \undefined \def \showISBNxiii  #1{\unskip}     \fi
\ifx \showISSN     \undefined \def \showISSN      #1{\unskip}     \fi
\ifx \showLCCN     \undefined \def \showLCCN      #1{\unskip}     \fi
\ifx \shownote     \undefined \def \shownote      #1{#1}          \fi
\ifx \showarticletitle \undefined \def \showarticletitle #1{#1}   \fi
\ifx \showURL      \undefined \def \showURL       {\relax}        \fi
\providecommand\bibfield[2]{#2}
\providecommand\bibinfo[2]{#2}
\providecommand\natexlab[1]{#1}
\providecommand\showeprint[2][]{arXiv:#2}

\bibitem[\protect\citeauthoryear{Asudeh, Jagadish, Stoyanovich, and Das}{Asudeh
  et~al\mbox{.}}{2019}]%
        {DBLP:conf/sigmod/AsudehJS019}
\bibfield{author}{\bibinfo{person}{Abolfazl Asudeh}, \bibinfo{person}{H.~V.
  Jagadish}, \bibinfo{person}{Julia Stoyanovich}, {and} \bibinfo{person}{Gautam
  Das}.} \bibinfo{year}{2019}\natexlab{}.
\newblock \showarticletitle{Designing Fair Ranking Schemes}. In
  \bibinfo{booktitle}{\emph{SIGMOD}}. \bibinfo{pages}{1259--1276}.
\newblock


\bibitem[\protect\citeauthoryear{Backes, Berrang, Humbert, and
  Manoharan}{Backes et~al\mbox{.}}{2016}]%
        {DBLP:conf/ccs/0001BHM16}
\bibfield{author}{\bibinfo{person}{Michael Backes}, \bibinfo{person}{Pascal
  Berrang}, \bibinfo{person}{Mathias Humbert}, {and} \bibinfo{person}{Praveen
  Manoharan}.} \bibinfo{year}{2016}\natexlab{}.
\newblock \showarticletitle{Membership Privacy in MicroRNA-based Studies}. In
  \bibinfo{booktitle}{\emph{CCS}}. \bibinfo{pages}{319--330}.
\newblock


\bibitem[\protect\citeauthoryear{Beigi, Mosallanezhad, Guo, Alvari, Nou, and
  Liu}{Beigi et~al\mbox{.}}{2020}]%
        {DBLP:conf/wsdm/BeigiMGAN020}
\bibfield{author}{\bibinfo{person}{Ghazaleh Beigi}, \bibinfo{person}{Ahmadreza
  Mosallanezhad}, \bibinfo{person}{Ruocheng Guo}, \bibinfo{person}{Hamidreza
  Alvari}, \bibinfo{person}{Alexander Nou}, {and} \bibinfo{person}{Huan Liu}.}
  \bibinfo{year}{2020}\natexlab{}.
\newblock \showarticletitle{Privacy-Aware Recommendation with Private-Attribute
  Protection using Adversarial Learning}. In \bibinfo{booktitle}{\emph{WSDM}}.
  \bibinfo{pages}{34--42}.
\newblock


\bibitem[\protect\citeauthoryear{Carlini, Liu, Erlingsson, Kos, and
  Song}{Carlini et~al\mbox{.}}{2019}]%
        {DBLP:conf/uss/Carlini0EKS19}
\bibfield{author}{\bibinfo{person}{Nicholas Carlini}, \bibinfo{person}{Chang
  Liu}, \bibinfo{person}{{\'{U}}lfar Erlingsson}, \bibinfo{person}{Jernej Kos},
  {and} \bibinfo{person}{Dawn Song}.} \bibinfo{year}{2019}\natexlab{}.
\newblock \showarticletitle{The Secret Sharer: Evaluating and Testing
  Unintended Memorization in Neural Networks}. In
  \bibinfo{booktitle}{\emph{USENIX}}. \bibinfo{pages}{267--284}.
\newblock


\bibitem[\protect\citeauthoryear{Chen, Yu, Zhang, and Fritz}{Chen
  et~al\mbox{.}}{2020b}]%
        {DBLP:conf/ccs/ChenYZF20}
\bibfield{author}{\bibinfo{person}{Dingfan Chen}, \bibinfo{person}{Ning Yu},
  \bibinfo{person}{Yang Zhang}, {and} \bibinfo{person}{Mario Fritz}.}
  \bibinfo{year}{2020}\natexlab{b}.
\newblock \showarticletitle{GAN-Leaks: {A} Taxonomy of Membership Inference
  Attacks against Generative Models}. In \bibinfo{booktitle}{\emph{CCS}}.
  \bibinfo{pages}{343--362}.
\newblock


\bibitem[\protect\citeauthoryear{Chen, Dong, Wang, Feng, Wang, and He}{Chen
  et~al\mbox{.}}{2020a}]%
        {DBLP:journals/corr/abs-2010-03240}
\bibfield{author}{\bibinfo{person}{Jiawei Chen}, \bibinfo{person}{Hande Dong},
  \bibinfo{person}{Xiang Wang}, \bibinfo{person}{Fuli Feng},
  \bibinfo{person}{Meng Wang}, {and} \bibinfo{person}{Xiangnan He}.}
  \bibinfo{year}{2020}\natexlab{a}.
\newblock \showarticletitle{Bias and Debias in Recommender System: {A} Survey
  and Future Directions}.
\newblock \bibinfo{journal}{\emph{CoRR}}  \bibinfo{volume}{abs/2010.03240}
  (\bibinfo{year}{2020}).
\newblock


\bibitem[\protect\citeauthoryear{Chen, Tang, Livescu, and Gimpel}{Chen
  et~al\mbox{.}}{2018}]%
        {DBLP:conf/emnlp/ChenTLG18}
\bibfield{author}{\bibinfo{person}{Mingda Chen}, \bibinfo{person}{Qingming
  Tang}, \bibinfo{person}{Karen Livescu}, {and} \bibinfo{person}{Kevin
  Gimpel}.} \bibinfo{year}{2018}\natexlab{}.
\newblock \showarticletitle{Variational Sequential Labelers for Semi-Supervised
  Learning}. In \bibinfo{booktitle}{\emph{EMNLP}}. \bibinfo{pages}{215--226}.
\newblock


\bibitem[\protect\citeauthoryear{Chen, Tang, Wiseman, and Gimpel}{Chen
  et~al\mbox{.}}{2019}]%
        {DBLP:conf/naacl/ChenTWG19}
\bibfield{author}{\bibinfo{person}{Mingda Chen}, \bibinfo{person}{Qingming
  Tang}, \bibinfo{person}{Sam Wiseman}, {and} \bibinfo{person}{Kevin Gimpel}.}
  \bibinfo{year}{2019}\natexlab{}.
\newblock \showarticletitle{A Multi-Task Approach for Disentangling Syntax and
  Semantics in Sentence Representations}. In
  \bibinfo{booktitle}{\emph{NAACL-HLT}}. \bibinfo{pages}{2453--2464}.
\newblock


\bibitem[\protect\citeauthoryear{Choquette{-}Choo, Tram{\`{e}}r, Carlini, and
  Papernot}{Choquette{-}Choo et~al\mbox{.}}{2021}]%
        {DBLP:conf/icml/Choquette-ChooT21}
\bibfield{author}{\bibinfo{person}{Christopher~A. Choquette{-}Choo},
  \bibinfo{person}{Florian Tram{\`{e}}r}, \bibinfo{person}{Nicholas Carlini},
  {and} \bibinfo{person}{Nicolas Papernot}.} \bibinfo{year}{2021}\natexlab{}.
\newblock \showarticletitle{Label-Only Membership Inference Attacks}. In
  \bibinfo{booktitle}{\emph{ICML}}, Vol.~\bibinfo{volume}{139}.
  \bibinfo{pages}{1964--1974}.
\newblock


\bibitem[\protect\citeauthoryear{Chu, Kim, and Han}{Chu et~al\mbox{.}}{2021}]%
        {chu2021learning}
\bibfield{author}{\bibinfo{person}{Sanghyeok Chu}, \bibinfo{person}{Dongwan
  Kim}, {and} \bibinfo{person}{Bohyung Han}.} \bibinfo{year}{2021}\natexlab{}.
\newblock \showarticletitle{Learning Debiased and Disentangled Representations
  for Semantic Segmentation}. In \bibinfo{booktitle}{\emph{NeurIPS}}.
  \bibinfo{pages}{8355--8366}.
\newblock


\bibitem[\protect\citeauthoryear{Davidson, Falorsi, Cao, Kipf, and
  Tomczak}{Davidson et~al\mbox{.}}{2018}]%
        {DBLP:conf/uai/DavidsonFCKT18}
\bibfield{author}{\bibinfo{person}{Tim~R. Davidson}, \bibinfo{person}{Luca
  Falorsi}, \bibinfo{person}{Nicola~De Cao}, \bibinfo{person}{Thomas Kipf},
  {and} \bibinfo{person}{Jakub~M. Tomczak}.} \bibinfo{year}{2018}\natexlab{}.
\newblock \showarticletitle{Hyperspherical Variational Auto-Encoders}. In
  \bibinfo{booktitle}{\emph{UAI}}. \bibinfo{pages}{856--865}.
\newblock


\bibitem[\protect\citeauthoryear{Devlin, Chang, Lee, and Toutanova}{Devlin
  et~al\mbox{.}}{2019}]%
        {DBLP:conf/naacl/DevlinCLT19}
\bibfield{author}{\bibinfo{person}{Jacob Devlin}, \bibinfo{person}{Ming{-}Wei
  Chang}, \bibinfo{person}{Kenton Lee}, {and} \bibinfo{person}{Kristina
  Toutanova}.} \bibinfo{year}{2019}\natexlab{}.
\newblock \showarticletitle{{BERT:} Pre-training of Deep Bidirectional
  Transformers for Language Understanding}. In
  \bibinfo{booktitle}{\emph{NAACL-HLT}}. \bibinfo{pages}{4171--4186}.
\newblock


\bibitem[\protect\citeauthoryear{Hagestedt, Zhang, Humbert, Berrang, Tang,
  Wang, and Backes}{Hagestedt et~al\mbox{.}}{2019}]%
        {DBLP:conf/ndss/Hagestedt0HBT0019}
\bibfield{author}{\bibinfo{person}{Inken Hagestedt}, \bibinfo{person}{Yang
  Zhang}, \bibinfo{person}{Mathias Humbert}, \bibinfo{person}{Pascal Berrang},
  \bibinfo{person}{Haixu Tang}, \bibinfo{person}{XiaoFeng Wang}, {and}
  \bibinfo{person}{Michael Backes}.} \bibinfo{year}{2019}\natexlab{}.
\newblock \showarticletitle{MBeacon: Privacy-Preserving Beacons for {DNA}
  Methylation Data}. In \bibinfo{booktitle}{\emph{NDSS}}.
\newblock


\bibitem[\protect\citeauthoryear{Harper and Konstan}{Harper and
  Konstan}{2016}]%
        {DBLP:journals/tiis/HarperK16}
\bibfield{author}{\bibinfo{person}{F.~Maxwell Harper} {and}
  \bibinfo{person}{Joseph~A. Konstan}.} \bibinfo{year}{2016}\natexlab{}.
\newblock \showarticletitle{The MovieLens Datasets: History and Context}.
\newblock \bibinfo{journal}{\emph{{ACM} Trans. Interact. Intell. Syst.}}
  \bibinfo{volume}{5}, \bibinfo{number}{4} (\bibinfo{year}{2016}),
  \bibinfo{pages}{19:1--19:19}.
\newblock


\bibitem[\protect\citeauthoryear{He, Liao, Zhang, Nie, Hu, and Chua}{He
  et~al\mbox{.}}{2017}]%
        {he2017neural}
\bibfield{author}{\bibinfo{person}{Xiangnan He}, \bibinfo{person}{Lizi Liao},
  \bibinfo{person}{Hanwang Zhang}, \bibinfo{person}{Liqiang Nie},
  \bibinfo{person}{Xia Hu}, {and} \bibinfo{person}{Tat-Seng Chua}.}
  \bibinfo{year}{2017}\natexlab{}.
\newblock \showarticletitle{Neural collaborative filtering}. In
  \bibinfo{booktitle}{\emph{WWW}}. \bibinfo{pages}{173--182}.
\newblock


\bibitem[\protect\citeauthoryear{Hidasi, Karatzoglou, Baltrunas, and
  Tikk}{Hidasi et~al\mbox{.}}{2016}]%
        {DBLP:journals/corr/HidasiKBT15}
\bibfield{author}{\bibinfo{person}{Bal{\'{a}}zs Hidasi},
  \bibinfo{person}{Alexandros Karatzoglou}, \bibinfo{person}{Linas Baltrunas},
  {and} \bibinfo{person}{Domonkos Tikk}.} \bibinfo{year}{2016}\natexlab{}.
\newblock \showarticletitle{Session-based Recommendations with Recurrent Neural
  Networks}. In \bibinfo{booktitle}{\emph{ICLR}}.
\newblock


\bibitem[\protect\citeauthoryear{Higgins, Matthey, Pal, Burgess, Glorot,
  Botvinick, Mohamed, and Lerchner}{Higgins et~al\mbox{.}}{2017}]%
        {DBLP:conf/iclr/HigginsMPBGBML17}
\bibfield{author}{\bibinfo{person}{Irina Higgins}, \bibinfo{person}{Lo{\"{\i}}c
  Matthey}, \bibinfo{person}{Arka Pal}, \bibinfo{person}{Christopher Burgess},
  \bibinfo{person}{Xavier Glorot}, \bibinfo{person}{Matthew Botvinick},
  \bibinfo{person}{Shakir Mohamed}, {and} \bibinfo{person}{Alexander
  Lerchner}.} \bibinfo{year}{2017}\natexlab{}.
\newblock \showarticletitle{beta-VAE: Learning Basic Visual Concepts with a
  Constrained Variational Framework}. In \bibinfo{booktitle}{\emph{ICLR}}.
\newblock


\bibitem[\protect\citeauthoryear{Homer, Szelinger, Redman, Duggan, Tembe,
  Muehling, Pearson, Stephan, Nelson, and Craig}{Homer et~al\mbox{.}}{2008}]%
        {homer2008resolving}
\bibfield{author}{\bibinfo{person}{Nils Homer}, \bibinfo{person}{Szabolcs
  Szelinger}, \bibinfo{person}{Margot Redman}, \bibinfo{person}{David Duggan},
  \bibinfo{person}{Waibhav Tembe}, \bibinfo{person}{Jill Muehling},
  \bibinfo{person}{John~V Pearson}, \bibinfo{person}{Dietrich~A Stephan},
  \bibinfo{person}{Stanley~F Nelson}, {and} \bibinfo{person}{David~W Craig}.}
  \bibinfo{year}{2008}\natexlab{}.
\newblock \showarticletitle{Resolving individuals contributing trace amounts of
  DNA to highly complex mixtures using high-density SNP genotyping
  microarrays}.
\newblock \bibinfo{journal}{\emph{PLoS genetics.}} \bibinfo{volume}{4},
  \bibinfo{number}{8} (\bibinfo{year}{2008}), \bibinfo{pages}{e1000167}.
\newblock


\bibitem[\protect\citeauthoryear{Jia, Salem, Backes, Zhang, and Gong}{Jia
  et~al\mbox{.}}{2019}]%
        {DBLP:conf/ccs/JiaSBZG19}
\bibfield{author}{\bibinfo{person}{Jinyuan Jia}, \bibinfo{person}{Ahmed Salem},
  \bibinfo{person}{Michael Backes}, \bibinfo{person}{Yang Zhang}, {and}
  \bibinfo{person}{Neil~Zhenqiang Gong}.} \bibinfo{year}{2019}\natexlab{}.
\newblock \showarticletitle{MemGuard: Defending against Black-Box Membership
  Inference Attacks via Adversarial Examples}. In
  \bibinfo{booktitle}{\emph{CCS}}. \bibinfo{pages}{259--274}.
\newblock


\bibitem[\protect\citeauthoryear{Kabbur, Ning, and Karypis}{Kabbur
  et~al\mbox{.}}{2013}]%
        {DBLP:conf/kdd/KabburNK13}
\bibfield{author}{\bibinfo{person}{Santosh Kabbur}, \bibinfo{person}{Xia Ning},
  {and} \bibinfo{person}{George Karypis}.} \bibinfo{year}{2013}\natexlab{}.
\newblock \showarticletitle{{FISM:} factored item similarity models for top-N
  recommender systems}. In \bibinfo{booktitle}{\emph{KDD}}.
  \bibinfo{pages}{659--667}.
\newblock


\bibitem[\protect\citeauthoryear{Kang, Fang, Wang, and McAuley}{Kang
  et~al\mbox{.}}{2017}]%
        {DBLP:conf/icdm/KangFWM17}
\bibfield{author}{\bibinfo{person}{Wang{-}Cheng Kang}, \bibinfo{person}{Chen
  Fang}, \bibinfo{person}{Zhaowen Wang}, {and} \bibinfo{person}{Julian~J.
  McAuley}.} \bibinfo{year}{2017}\natexlab{}.
\newblock \showarticletitle{Visually-Aware Fashion Recommendation and Design
  with Generative Image Models}. In \bibinfo{booktitle}{\emph{ICDM}}.
  \bibinfo{pages}{207--216}.
\newblock


\bibitem[\protect\citeauthoryear{Kang and McAuley}{Kang and McAuley}{2018}]%
        {DBLP:conf/icdm/KangM18}
\bibfield{author}{\bibinfo{person}{Wang{-}Cheng Kang} {and}
  \bibinfo{person}{Julian~J. McAuley}.} \bibinfo{year}{2018}\natexlab{}.
\newblock \showarticletitle{Self-Attentive Sequential Recommendation}. In
  \bibinfo{booktitle}{\emph{ICDM}}. \bibinfo{pages}{197--206}.
\newblock


\bibitem[\protect\citeauthoryear{Kim, Park, Oh, Lee, and Yu}{Kim
  et~al\mbox{.}}{2016}]%
        {DBLP:conf/recsys/KimPOLY16}
\bibfield{author}{\bibinfo{person}{Dong~Hyun Kim}, \bibinfo{person}{Chanyoung
  Park}, \bibinfo{person}{Jinoh Oh}, \bibinfo{person}{Sungyoung Lee}, {and}
  \bibinfo{person}{Hwanjo Yu}.} \bibinfo{year}{2016}\natexlab{}.
\newblock \showarticletitle{Convolutional Matrix Factorization for Document
  Context-Aware Recommendation}. In \bibinfo{booktitle}{\emph{RecSys}}.
  \bibinfo{pages}{233--240}.
\newblock


\bibitem[\protect\citeauthoryear{Kim and Mnih}{Kim and Mnih}{2018}]%
        {DBLP:conf/icml/KimM18}
\bibfield{author}{\bibinfo{person}{Hyunjik Kim} {and} \bibinfo{person}{Andriy
  Mnih}.} \bibinfo{year}{2018}\natexlab{}.
\newblock \showarticletitle{Disentangling by Factorising}. In
  \bibinfo{booktitle}{\emph{ICML}}, Vol.~\bibinfo{volume}{80}.
  \bibinfo{pages}{2654--2663}.
\newblock


\bibitem[\protect\citeauthoryear{Koren}{Koren}{2008}]%
        {DBLP:conf/kdd/Koren08}
\bibfield{author}{\bibinfo{person}{Yehuda Koren}.}
  \bibinfo{year}{2008}\natexlab{}.
\newblock \showarticletitle{Factorization meets the neighborhood: a
  multifaceted collaborative filtering model}. In
  \bibinfo{booktitle}{\emph{KDD}}. \bibinfo{pages}{426--434}.
\newblock


\bibitem[\protect\citeauthoryear{Koren and Bell}{Koren and Bell}{2015}]%
        {DBLP:reference/sp/KorenB15}
\bibfield{author}{\bibinfo{person}{Yehuda Koren} {and}
  \bibinfo{person}{Robert~M. Bell}.} \bibinfo{year}{2015}\natexlab{}.
\newblock \showarticletitle{Advances in Collaborative Filtering}.
\newblock In \bibinfo{booktitle}{\emph{Recommender Systems Handbook}}.
  \bibinfo{pages}{77--118}.
\newblock


\bibitem[\protect\citeauthoryear{Koren, Bell, and Volinsky}{Koren
  et~al\mbox{.}}{2009}]%
        {DBLP:journals/computer/KorenBV09}
\bibfield{author}{\bibinfo{person}{Yehuda Koren}, \bibinfo{person}{Robert~M.
  Bell}, {and} \bibinfo{person}{Chris Volinsky}.}
  \bibinfo{year}{2009}\natexlab{}.
\newblock \showarticletitle{Matrix Factorization Techniques for Recommender
  Systems}.
\newblock \bibinfo{journal}{\emph{Computer.}} \bibinfo{volume}{42},
  \bibinfo{number}{8} (\bibinfo{year}{2009}), \bibinfo{pages}{30--37}.
\newblock


\bibitem[\protect\citeauthoryear{Kusner, Loftus, Russell, and Silva}{Kusner
  et~al\mbox{.}}{2017}]%
        {DBLP:conf/nips/KusnerLRS17}
\bibfield{author}{\bibinfo{person}{Matt~J. Kusner}, \bibinfo{person}{Joshua~R.
  Loftus}, \bibinfo{person}{Chris Russell}, {and} \bibinfo{person}{Ricardo
  Silva}.} \bibinfo{year}{2017}\natexlab{}.
\newblock \showarticletitle{Counterfactual Fairness}. In
  \bibinfo{booktitle}{\emph{NeurIPS}}. \bibinfo{pages}{4066--4076}.
\newblock


\bibitem[\protect\citeauthoryear{Li and Zhang}{Li and Zhang}{2021}]%
        {DBLP:conf/ccs/LiZ21}
\bibfield{author}{\bibinfo{person}{Zheng Li} {and} \bibinfo{person}{Yang
  Zhang}.} \bibinfo{year}{2021}\natexlab{}.
\newblock \showarticletitle{Membership Leakage in Label-Only Exposures}. In
  \bibinfo{booktitle}{\emph{CCS}}. \bibinfo{pages}{880--895}.
\newblock


\bibitem[\protect\citeauthoryear{Lin, Sonboli, Mobasher, and Burke}{Lin
  et~al\mbox{.}}{2019}]%
        {DBLP:conf/recsys/LinSMB19}
\bibfield{author}{\bibinfo{person}{Kun Lin}, \bibinfo{person}{Nasim Sonboli},
  \bibinfo{person}{Bamshad Mobasher}, {and} \bibinfo{person}{Robin Burke}.}
  \bibinfo{year}{2019}\natexlab{}.
\newblock \showarticletitle{Crank up the Volume: Preference Bias Amplification
  in Collaborative Recommendation}. In \bibinfo{booktitle}{\emph{RecSys}},
  Vol.~\bibinfo{volume}{2440}.
\newblock


\bibitem[\protect\citeauthoryear{Linden, Smith, and York}{Linden
  et~al\mbox{.}}{2003}]%
        {DBLP:journals/internet/LindenSY03}
\bibfield{author}{\bibinfo{person}{Greg Linden}, \bibinfo{person}{Brent Smith},
  {and} \bibinfo{person}{Jeremy York}.} \bibinfo{year}{2003}\natexlab{}.
\newblock \showarticletitle{Amazon.com Recommendations: Item-to-Item
  Collaborative Filtering}.
\newblock \bibinfo{journal}{\emph{{IEEE} Internet Comput.}}
  \bibinfo{volume}{7}, \bibinfo{number}{1} (\bibinfo{year}{2003}),
  \bibinfo{pages}{76--80}.
\newblock


\bibitem[\protect\citeauthoryear{Marlin, Zemel, Roweis, and Slaney}{Marlin
  et~al\mbox{.}}{2007}]%
        {DBLP:conf/uai/MarlinZRS07}
\bibfield{author}{\bibinfo{person}{Benjamin~M. Marlin},
  \bibinfo{person}{Richard~S. Zemel}, \bibinfo{person}{Sam~T. Roweis}, {and}
  \bibinfo{person}{Malcolm Slaney}.} \bibinfo{year}{2007}\natexlab{}.
\newblock \showarticletitle{Collaborative Filtering and the Missing at Random
  Assumption}. In \bibinfo{booktitle}{\emph{UAI}}. \bibinfo{pages}{267--275}.
\newblock


\bibitem[\protect\citeauthoryear{McAuley, Targett, Shi, and van~den
  Hengel}{McAuley et~al\mbox{.}}{2015}]%
        {DBLP:conf/sigir/McAuleyTSH15}
\bibfield{author}{\bibinfo{person}{Julian~J. McAuley},
  \bibinfo{person}{Christopher Targett}, \bibinfo{person}{Qinfeng Shi}, {and}
  \bibinfo{person}{Anton van~den Hengel}.} \bibinfo{year}{2015}\natexlab{}.
\newblock \showarticletitle{Image-Based Recommendations on Styles and
  Substitutes}. In \bibinfo{booktitle}{\emph{SIGIR}}. \bibinfo{pages}{43--52}.
\newblock


\bibitem[\protect\citeauthoryear{Nasr, Shokri, and Houmansadr}{Nasr
  et~al\mbox{.}}{2018}]%
        {DBLP:conf/ccs/NasrSH18}
\bibfield{author}{\bibinfo{person}{Milad Nasr}, \bibinfo{person}{Reza Shokri},
  {and} \bibinfo{person}{Amir Houmansadr}.} \bibinfo{year}{2018}\natexlab{}.
\newblock \showarticletitle{Machine Learning with Membership Privacy using
  Adversarial Regularization}. In \bibinfo{booktitle}{\emph{CCS}}.
  \bibinfo{pages}{634--646}.
\newblock


\bibitem[\protect\citeauthoryear{Nasr, Shokri, and Houmansadr}{Nasr
  et~al\mbox{.}}{2019}]%
        {DBLP:conf/sp/NasrSH19}
\bibfield{author}{\bibinfo{person}{Milad Nasr}, \bibinfo{person}{Reza Shokri},
  {and} \bibinfo{person}{Amir Houmansadr}.} \bibinfo{year}{2019}\natexlab{}.
\newblock \showarticletitle{{Comprehensive Privacy Analysis of Deep Learning:
  Passive and Active White-box Inference Attacks against Centralized and
  Federated Learning}}. In \bibinfo{booktitle}{\emph{SP}}.
  \bibinfo{pages}{1021--1035}.
\newblock


\bibitem[\protect\citeauthoryear{Polat and Du}{Polat and Du}{2005}]%
        {DBLP:conf/sac/PolatD05}
\bibfield{author}{\bibinfo{person}{Huseyin Polat} {and}
  \bibinfo{person}{Wenliang Du}.} \bibinfo{year}{2005}\natexlab{}.
\newblock \showarticletitle{SVD-based collaborative filtering with privacy}. In
  \bibinfo{booktitle}{\emph{SAC}}. \bibinfo{pages}{791--795}.
\newblock


\bibitem[\protect\citeauthoryear{Pyrgelis, Troncoso, and Cristofaro}{Pyrgelis
  et~al\mbox{.}}{2018}]%
        {DBLP:conf/ndss/PyrgelisTC18}
\bibfield{author}{\bibinfo{person}{Apostolos Pyrgelis},
  \bibinfo{person}{Carmela Troncoso}, {and} \bibinfo{person}{Emiliano~De
  Cristofaro}.} \bibinfo{year}{2018}\natexlab{}.
\newblock \showarticletitle{Knock Knock, Who's There? Membership Inference on
  Aggregate Location Data}. In \bibinfo{booktitle}{\emph{NDSS}}.
\newblock


\bibitem[\protect\citeauthoryear{Qian, Feng, Wen, and Chua}{Qian
  et~al\mbox{.}}{2021}]%
        {DBLP:conf/aaai/QianFWC21}
\bibfield{author}{\bibinfo{person}{Chen Qian}, \bibinfo{person}{Fuli Feng},
  \bibinfo{person}{Lijie Wen}, {and} \bibinfo{person}{Tat{-}Seng Chua}.}
  \bibinfo{year}{2021}\natexlab{}.
\newblock \showarticletitle{Conceptualized and Contextualized Gaussian
  Embedding}. In \bibinfo{booktitle}{\emph{AAAI}}.
  \bibinfo{pages}{13683--13691}.
\newblock


\bibitem[\protect\citeauthoryear{Rendle, Freudenthaler, and
  Schmidt{-}Thieme}{Rendle et~al\mbox{.}}{2010}]%
        {DBLP:conf/www/RendleFS10}
\bibfield{author}{\bibinfo{person}{Steffen Rendle}, \bibinfo{person}{Christoph
  Freudenthaler}, {and} \bibinfo{person}{Lars Schmidt{-}Thieme}.}
  \bibinfo{year}{2010}\natexlab{}.
\newblock \showarticletitle{Factorizing personalized Markov chains for
  next-basket recommendation}. In \bibinfo{booktitle}{\emph{WWW}}.
  \bibinfo{pages}{811--820}.
\newblock


\bibitem[\protect\citeauthoryear{Saito}{Saito}{2020}]%
        {DBLP:conf/sigir/Saito20}
\bibfield{author}{\bibinfo{person}{Yuta Saito}.}
  \bibinfo{year}{2020}\natexlab{}.
\newblock \showarticletitle{Asymmetric Tri-training for Debiasing
  Missing-Not-At-Random Explicit Feedback}. In
  \bibinfo{booktitle}{\emph{SIGIR}}. \bibinfo{pages}{309--318}.
\newblock


\bibitem[\protect\citeauthoryear{Salakhutdinov and Mnih}{Salakhutdinov and
  Mnih}{2007}]%
        {mnih2007probabilistic}
\bibfield{author}{\bibinfo{person}{Ruslan Salakhutdinov} {and}
  \bibinfo{person}{Andriy Mnih}.} \bibinfo{year}{2007}\natexlab{}.
\newblock \showarticletitle{Probabilistic Matrix Factorization}. In
  \bibinfo{booktitle}{\emph{NeurIPS}}. \bibinfo{pages}{1257--1264}.
\newblock


\bibitem[\protect\citeauthoryear{Salem, Zhang, Humbert, Berrang, Fritz, and
  Backes}{Salem et~al\mbox{.}}{2019}]%
        {DBLP:conf/ndss/Salem0HBF019}
\bibfield{author}{\bibinfo{person}{Ahmed Salem}, \bibinfo{person}{Yang Zhang},
  \bibinfo{person}{Mathias Humbert}, \bibinfo{person}{Pascal Berrang},
  \bibinfo{person}{Mario Fritz}, {and} \bibinfo{person}{Michael Backes}.}
  \bibinfo{year}{2019}\natexlab{}.
\newblock \showarticletitle{ML-Leaks: Model and Data Independent Membership
  Inference Attacks and Defenses on Machine Learning Models}. In
  \bibinfo{booktitle}{\emph{NDSS}}.
\newblock


\bibitem[\protect\citeauthoryear{Sarwar, Karypis, Konstan, and Riedl}{Sarwar
  et~al\mbox{.}}{2001}]%
        {DBLP:conf/www/SarwarKKR01}
\bibfield{author}{\bibinfo{person}{Badrul~Munir Sarwar},
  \bibinfo{person}{George Karypis}, \bibinfo{person}{Joseph~A. Konstan}, {and}
  \bibinfo{person}{John Riedl}.} \bibinfo{year}{2001}\natexlab{}.
\newblock \showarticletitle{Item-based collaborative filtering recommendation
  algorithms}. In \bibinfo{booktitle}{\emph{WWW}}. \bibinfo{pages}{285--295}.
\newblock


\bibitem[\protect\citeauthoryear{Schnabel, Swaminathan, Singh, Chandak, and
  Joachims}{Schnabel et~al\mbox{.}}{2016}]%
        {DBLP:conf/icml/SchnabelSSCJ16}
\bibfield{author}{\bibinfo{person}{Tobias Schnabel}, \bibinfo{person}{Adith
  Swaminathan}, \bibinfo{person}{Ashudeep Singh}, \bibinfo{person}{Navin
  Chandak}, {and} \bibinfo{person}{Thorsten Joachims}.}
  \bibinfo{year}{2016}\natexlab{}.
\newblock \showarticletitle{Recommendations as Treatments: Debiasing Learning
  and Evaluation}. In \bibinfo{booktitle}{\emph{ICML}},
  Vol.~\bibinfo{volume}{48}. \bibinfo{pages}{1670--1679}.
\newblock


\bibitem[\protect\citeauthoryear{Sedhain, Menon, Sanner, and Xie}{Sedhain
  et~al\mbox{.}}{2015}]%
        {DBLP:conf/www/SedhainMSX15}
\bibfield{author}{\bibinfo{person}{Suvash Sedhain},
  \bibinfo{person}{Aditya~Krishna Menon}, \bibinfo{person}{Scott Sanner}, {and}
  \bibinfo{person}{Lexing Xie}.} \bibinfo{year}{2015}\natexlab{}.
\newblock \showarticletitle{AutoRec: Autoencoders Meet Collaborative
  Filtering}. In \bibinfo{booktitle}{\emph{WWW}}. \bibinfo{pages}{111--112}.
\newblock


\bibitem[\protect\citeauthoryear{Shokri, Stronati, Song, and Shmatikov}{Shokri
  et~al\mbox{.}}{2017}]%
        {DBLP:conf/sp/ShokriSSS17}
\bibfield{author}{\bibinfo{person}{Reza Shokri}, \bibinfo{person}{Marco
  Stronati}, \bibinfo{person}{Congzheng Song}, {and} \bibinfo{person}{Vitaly
  Shmatikov}.} \bibinfo{year}{2017}\natexlab{}.
\newblock \showarticletitle{Membership Inference Attacks Against Machine
  Learning Models}. In \bibinfo{booktitle}{\emph{SP}}. \bibinfo{pages}{3--18}.
\newblock


\bibitem[\protect\citeauthoryear{Steck}{Steck}{2010}]%
        {DBLP:conf/kdd/Steck10}
\bibfield{author}{\bibinfo{person}{Harald Steck}.}
  \bibinfo{year}{2010}\natexlab{}.
\newblock \showarticletitle{Training and testing of recommender systems on data
  missing not at random}. In \bibinfo{booktitle}{\emph{KDD}}.
  \bibinfo{pages}{713--722}.
\newblock


\bibitem[\protect\citeauthoryear{Sun, Liu, Wu, Pei, Lin, Ou, and Jiang}{Sun
  et~al\mbox{.}}{2019}]%
        {DBLP:conf/cikm/SunLWPLOJ19}
\bibfield{author}{\bibinfo{person}{Fei Sun}, \bibinfo{person}{Jun Liu},
  \bibinfo{person}{Jian Wu}, \bibinfo{person}{Changhua Pei},
  \bibinfo{person}{Xiao Lin}, \bibinfo{person}{Wenwu Ou}, {and}
  \bibinfo{person}{Peng Jiang}.} \bibinfo{year}{2019}\natexlab{}.
\newblock \showarticletitle{BERT4Rec: Sequential Recommendation with
  Bidirectional Encoder Representations from Transformer}. In
  \bibinfo{booktitle}{\emph{CIKM}}. \bibinfo{pages}{1441--1450}.
\newblock


\bibitem[\protect\citeauthoryear{Tang and Wang}{Tang and Wang}{2018}]%
        {DBLP:conf/wsdm/TangW18}
\bibfield{author}{\bibinfo{person}{Jiaxi Tang} {and} \bibinfo{person}{Ke
  Wang}.} \bibinfo{year}{2018}\natexlab{}.
\newblock \showarticletitle{Personalized Top-N Sequential Recommendation via
  Convolutional Sequence Embedding}. In \bibinfo{booktitle}{\emph{WSDM}}.
  \bibinfo{pages}{565--573}.
\newblock


\bibitem[\protect\citeauthoryear{van~der Maaten and Hinton}{van~der Maaten and
  Hinton}{2008}]%
        {JMLR:v9:vandermaaten08a}
\bibfield{author}{\bibinfo{person}{Laurens van~der Maaten} {and}
  \bibinfo{person}{Geoffrey Hinton}.} \bibinfo{year}{2008}\natexlab{}.
\newblock \showarticletitle{Visualizing Data using t-SNE}.
\newblock \bibinfo{journal}{\emph{J. Mach. Learn. Res.}} \bibinfo{volume}{9},
  \bibinfo{number}{86} (\bibinfo{year}{2008}), \bibinfo{pages}{2579--2605}.
\newblock


\bibitem[\protect\citeauthoryear{Wang, Zhang, Sun, and Qi}{Wang
  et~al\mbox{.}}{2021}]%
        {DBLP:conf/wsdm/0003ZS021}
\bibfield{author}{\bibinfo{person}{Xiaojie Wang}, \bibinfo{person}{Rui Zhang},
  \bibinfo{person}{Yu Sun}, {and} \bibinfo{person}{Jianzhong Qi}.}
  \bibinfo{year}{2021}\natexlab{}.
\newblock \showarticletitle{Combating Selection Biases in Recommender Systems
  with a Few Unbiased Ratings}. In \bibinfo{booktitle}{\emph{WSDM}}.
  \bibinfo{pages}{427--435}.
\newblock


\bibitem[\protect\citeauthoryear{Yeom, Giacomelli, Fredrikson, and Jha}{Yeom
  et~al\mbox{.}}{2018}]%
        {DBLP:conf/csfw/YeomGFJ18}
\bibfield{author}{\bibinfo{person}{Samuel Yeom}, \bibinfo{person}{Irene
  Giacomelli}, \bibinfo{person}{Matt Fredrikson}, {and} \bibinfo{person}{Somesh
  Jha}.} \bibinfo{year}{2018}\natexlab{}.
\newblock \showarticletitle{Privacy Risk in Machine Learning: Analyzing the
  Connection to Overfitting}. In \bibinfo{booktitle}{\emph{CSF}}.
  \bibinfo{pages}{268--282}.
\newblock


\bibitem[\protect\citeauthoryear{Zhang, Ren, Wang, Ren, Chen, Hu, and
  Zhang}{Zhang et~al\mbox{.}}{2021}]%
        {DBLP:conf/ccs/ZhangRWRCHZ21}
\bibfield{author}{\bibinfo{person}{Minxing Zhang}, \bibinfo{person}{Zhaochun
  Ren}, \bibinfo{person}{Zihan Wang}, \bibinfo{person}{Pengjie Ren},
  \bibinfo{person}{Zhumin Chen}, \bibinfo{person}{Pengfei Hu}, {and}
  \bibinfo{person}{Yang Zhang}.} \bibinfo{year}{2021}\natexlab{}.
\newblock \showarticletitle{Membership Inference Attacks Against Recommender
  Systems}. In \bibinfo{booktitle}{\emph{CCS}}. \bibinfo{pages}{864--879}.
\newblock


\bibitem[\protect\citeauthoryear{Zhou and Neubig}{Zhou and Neubig}{2017}]%
        {DBLP:conf/acl/ZhouN17}
\bibfield{author}{\bibinfo{person}{Chunting Zhou} {and} \bibinfo{person}{Graham
  Neubig}.} \bibinfo{year}{2017}\natexlab{}.
\newblock \showarticletitle{Multi-space Variational Encoder-Decoders for
  Semi-supervised Labeled Sequence Transduction}. In
  \bibinfo{booktitle}{\emph{ACL}}. \bibinfo{pages}{310--320}.
\newblock


\bibitem[\protect\citeauthoryear{Zhou, Mou, Fan, Pi, Bian, Zhou, Zhu, and
  Gai}{Zhou et~al\mbox{.}}{2019}]%
        {DBLP:conf/aaai/ZhouMFPBZZG19}
\bibfield{author}{\bibinfo{person}{Guorui Zhou}, \bibinfo{person}{Na Mou},
  \bibinfo{person}{Ying Fan}, \bibinfo{person}{Qi Pi}, \bibinfo{person}{Weijie
  Bian}, \bibinfo{person}{Chang Zhou}, \bibinfo{person}{Xiaoqiang Zhu}, {and}
  \bibinfo{person}{Kun Gai}.} \bibinfo{year}{2019}\natexlab{}.
\newblock \showarticletitle{Deep Interest Evolution Network for Click-Through
  Rate Prediction}. In \bibinfo{booktitle}{\emph{AAAI}}.
  \bibinfo{pages}{5941--5948}.
\newblock


\end{thebibliography}

\clearpage
\appendix
\section{Appendix}

\subsection{Learning algorithm of \ac{DL-MIA}}
\label{subsec:learning algorithm and training}

Algorithm~\ref{algorithm:Training algorithm} gives the detailed learning algorithm of \ac{DL-MIA}.
Specifically, given the target recommender, we first establish the shadow recommender $\mathcal{M}_{\mathit{shadow}}$, calculate the difference vector $\mathbf{f}^{\mathit{diff}}$ by \ac{MF}, and initialize model parameter $\mathbf{\Theta}$ for the disentangled encoder and attack model (line 1--3).
Then, to mitigate the gap between the shadow and target recommenders, we train the disentangled encoder and attack model by jointly optimizing $\mathcal{L}_{\mathit{BCE}}$ and $\mathcal{L}_{\mathit{ELBO}}$ (line 4--7).
In this way, the disentangled difference vector $\mathbf{f}^{\mathit{dis}}$ is computed.
Next, to further reduce the influence of the estimation bias, an alternating training strategy is adopted.
By determining data sample weights $\mathbf{w}$ using the current $\mathbf{p}$, we are able to minimize the reweighted loss $\mathcal{L}_{\mathit{reweight}}$, and obtain the reweighted difference vector $\mathbf{f}^{\mathit{rew}}$ (line 10--15).
With input of $\mathbf{f}^{\mathit{rew}}$ and $y_{target}$, the current $p$ can be refined using the estimation constraint $\mathcal{L}_{estimation}$ (line 16--19).  
During the alternating training process, $\mathcal{L}_{\mathit{reweight}}$ and $\mathcal{L}_{\mathit{estimation}}$ are optimized iteratively (line 9--20).

\begin{algorithm}[!h]
    \caption{Training algorithm of \ac{DL-MIA}.}
    \label{algorithm:Training algorithm}      
    \begin{algorithmic}[1] 
    \Require The trained shadow recommender $\mathcal{M}_{\mathit{shadow}}$; the difference vector $\mathbf{f}^{\mathit{diff}}$ from the generator; randomly initialized truth-level scores $\mathbf{p}$; the number of inner-loop epochs $\mathit{epoch}_{\mathit{in}}$ and outer-loop epochs $\mathit{epoch}_{\mathit{out}}$ for the alternating training; the number of epochs for pretraining $\mathit{epoch}_{\mathit{pre}}$; parameters $\mathbf{\Theta}$ for the disentangled encoder and attack model.
    \Ensure The disentangled and reweighted difference vector $\mathbf{f}^{\mathit{rew}}$ and trained attack model $\mathcal{A}$;
    \State Establish the shadow recommender $\mathcal{M}_{\mathit{shadow}}$;
    \State Calculate the difference vector $\mathbf{f}^{\mathit{diff}}$ (Eq.~\ref{eq:difference vector generator});
    \State Randomly initialize model parameters $\mathbf{\Theta}$;
    \While{$i\leq\mathit{epoch}_{\mathit{pre}}$ }
    \State Calculate the disentangled difference vector $\mathbf{f}^{\mathit{dis}}$;
    \State Input $\mathbf{f}^{\mathit{dis}}$ into the attack model $\mathcal{A}$ for predicting $y_{shadow}$;
    \State Update $\mathbf{\Theta}$ by jointly optimizing $\mathcal{L}_{\mathit{BCE}}$ and $\mathcal{L}_{\mathit{ELBO}}$ (Eq.~\ref{eq:BCE} and~\ref{eq:ELBO});
    \EndWhile
    
    \While{$i\leq\mathit{epoch}_{\mathit{out}}$}
    \State Compute the data sample weights $\mathbf{w}$ using the current $\mathbf{p}$;
    \While{$j\leq\mathit{epoch}_{\mathit{in}}$}
    \State Calculate the reweighted feature vector $\mathbf{f}^{\mathit{rew}}$;
    \State Input $\mathbf{f}^{\mathit{rew}}$ into $\mathcal{A}$ for predicting $y_{target}$ and $y_{shadow}$;
    \State Update $\mathbf{\Theta}$ by minizing $\mathcal{L}_{\mathit{reweight}}$ (Eq.~\ref{eq:reweight loss});
    \EndWhile
    \State Input $\mathbf{f}^{\mathit{rew}}$ and $y_{target}$ into the weight estimator;
    \While{$k\leq\mathit{epoch}_{\mathit{in}}$}
    \State Refine the current truth-level scores $\mathbf{p}$ by optimizing $\mathcal{L}_{\mathit{estimation}}$ (Eq.~\ref{eq:truth-level score} and~\ref{eq:estimation constraint});
    \EndWhile
    \EndWhile
    \end{algorithmic}
    \end{algorithm}

\subsection{Notation}
\label{subsec:Notations for different settings}
Table~\ref{tab:Notations for different settings.} shows the notation we use for different experimental settings.

\subsection{Implementation details}
\label{subsec:implementation details for baselines}
Table~\ref{tab:Parameter settings of different recommender systems} demonstrate the parameter settings of different recommenders in our experiments. 
Note that, we do not give the parameter setting of  \ac{ItemBase}~\citep{DBLP:conf/www/SarwarKKR01} since it is based on the statistical method.

\subsection{Reproducibility}
To facilitate the reproducibility of the results reported in this work, the
code and data used in this work is available at \url{https://github.com/WZH-NLP/DL-MIA-KDD-2022}.

\begin{table}[t]
    \small
  \centering
  \setlength{\tabcolsep}{1mm}
   \caption{Notation for different settings.  ``Rec.'' is short for ``Recommender.'' ``$*$'' stands for any recommendation algorithm or dataset. }
  \begin{tabular}{lcp{6.25cm}}
  \toprule
   \bf Rec.& \bf Notation &  \bf Description \\ 
   \midrule
  \multirow{11}{*}{\rotatebox[origin=c]{90}{General}}
   &  \multirow{1}{*}{A$*$} & Trained on the Amazon Digital Music dataset. 
   \\ \cline{2-3}
   &  M$*$ & Trained on the MovieLens-1M dataset.
   \\\cline{2-3}
   &  $*$I & Implemented by the ItemBase algorithm.  
   \\\cline{2-3}
   &  $*$L & Implemented by the LFM algorithm. 
   \\\cline{2-3}  
   & $*$N &  Implemented by the NCF algorithm. 
   \\\cline{2-3}
   & \multirow{2}{*}{AI} & The recommender is implemented by Item algorithm on the Amazon Digital Music dataset. 
   \\\cline{2-3}
   &\multirow{4}{*} {AIMN} &
   The shadow recommender is implemented 
   by the ItemBase algorithm on the Amazon Digital 
   Music dataset, and the target recommender 
   is implemented by NCF algorithm on the
   MovieLens-1M dataset.\\     
   \midrule
   \multirow{11}{*}{\rotatebox[origin=c]{90}{Sequential}}
   & A$*$ & Trained on the Amazon Beauty datset.\\ \cline{2-3}
   & M$*$ & Trained on the MovieLens-1M dataset.\\ \cline{2-3}
   & $*$B & Implemented by BERT4REC algorithm.\\ \cline{2-3}
   & $*$C & Implemented by Caser algorithm.\\ \cline{2-3}
   & $*$G & Implemented by GRU4REC algorithm.\\ 
   \cline{2-3}
   & \multirow{2}{*}{AB} & The recommender is implemented 
   by BERT4REC on the Amazon Beauty dataset. \\ 
   \cline{2-3}
   &\multirow{4}{*}{ABMC} &The shadow recommender is implemented 
   by BERT4REC on the Amazon Beauty dataset, 
   and the target recommender is implemented 
   by Caser on the MovieLens-1M dataset.\\ 
  \bottomrule
  \label{tab:Notations for different settings.}
  \end{tabular}
  \end{table}

\begin{table}
    \small
	\centering
	\caption{Parameter settings of different recommender systems.}
	\begin{tabular}{l p{6.3cm}}
		\toprule
		\textbf{Baseline} & \textbf{Settings}\\
		\midrule
		\multirow{1}{*}{ItemBase}
		& -- \\ 
		\midrule
		LFM
		& Embed.-size=100, 
		SGD optimizer, 
		learning-rate=0.01 \\ 
		\midrule
		\multirow{2}{*}{NCF}
		& Embed.-size=8, batch-size=256,   
		Adam optimizer, 
		hidden-size=64, 32, 16, 
		learning-rate=0.001\\ 
		\midrule
		\multirow{2}{*}{BERT4REC}
		& batch-size=128,   
		Adam optimizer, dropout=0.1
		hidden-size=256, 
		learning-rate=0.001\\ 
		\midrule
		\multirow{2}{*}{Caser}
		& Embed.-size=50, batch-size=512,   
		Adam optimizer, dropout=0.5
		learning-rate=0.001\\ 
		\midrule
		\multirow{2}{*}{GRU4REC}
		& batch-size=50,  
		Adagrad optimizer, dropout=0.5, 
		hidden-size=100, 
		learning-rate=0.01, momentum=0\\
		\bottomrule
	\end{tabular}
	\label{tab:Parameter settings of different recommender systems}
\end{table}

\end{document}